\documentclass[longauth,letterpaper]{aa} 
\usepackage{graphicx}
\usepackage{natbib}
\usepackage{txfonts}
\usepackage{longtable}
\newcommand{\hess}{H\textsc{.E.S.S.}}
\newcommand{\gr}{\ensuremath{\gamma}-ray}
\newcommand{\grs}{\ensuremath{\gamma}-rays}
\newcommand{\dg}{\ensuremath{^{\circ}}}
\newcommand{\on}{\emph{on}}
\newcommand{\off}{\emph{off}}
\newcommand{\thsq}{\ensuremath{\theta^{2}}}
\newcommand{\dflux}{\ensuremath{\textrm{cm}^{-2} \textrm{s}^{-1} \textrm{TeV}^{-1}}}
\newcommand{\iflux}{\ensuremath{\textrm{cm}^{-2} \textrm{s}^{-1}}}
\newcommand{\pmin}{\ensuremath{\gamma\ \textrm{min}^{-1}}}
\newcommand{\atrue}{\ensuremath{\textrm{A}true}}
\newcommand{\areco}{\ensuremath{\textrm{A}reco}}
\newcommand{\chisq}{\ensuremath{\chi^{2}}}
\newcommand{\rchisq}{\ensuremath{\chi^{2}}/dof}
\newcommand{\stat}{\ensuremath{_{\textrm{\tiny{stat}}}}}
\newcommand{\sys}{\ensuremath{_{\textrm{\tiny{sys}}}}}
\newcommand{\sigrate}{\ensuremath{\sigma/\sqrt{t\,(\mathrm{hr}^{-1})}}}

\begin{document}

\title{Observations of the Crab Nebula with H.E.S.S.}

\author{F. Aharonian\inst{1}
 \and A.G.~Akhperjanian \inst{2}
 \and A.R.~Bazer-Bachi \inst{3}
 \and M.~Beilicke \inst{4}
 \and W.~Benbow \inst{1}
 \and D.~Berge \inst{1}
 \and K.~Bernl\"ohr \inst{1,5}
 \and C.~Boisson \inst{6}
 \and O.~Bolz \inst{1}
 \and V.~Borrel \inst{3}
% ...
}

\institute{
Max-Planck-Institut f\"ur Kernphysik, P.O. Box 103980, D 69029
Heidelberg, Germany
\and
 Yerevan Physics Institute, 2 Alikhanian Brothers St., 375036 Yerevan,
Armenia
\and
Centre d'Etude Spatiale des Rayonnements, CNRS/UPS, 9 av. du Colonel Roche, BP
4346, F-31029 Toulouse Cedex 4, France
\and
Universit\"at Hamburg, Institut f\"ur Experimentalphysik, Luruper Chaussee
149, D 22761 Hamburg, Germany
\and
Institut f\"ur Physik, Humboldt-Universit\"at zu Berlin, Newtonstr. 15,
D 12489 Berlin, Germany
\and
LUTH, UMR 8102 du CNRS, Observatoire de Paris, Section de Meudon, F-92195 Meudon Cedex,
France
% ...
}

\date{Received / Accepted }
  
\abstract{ The Crab nebula was observed with the \hess\ stereoscopic
  Cherenkov-telescope array between October 2003 and January 2005 for
  a total of 22.9 hours (after data quality selection). This period of
  time partly overlapped with the commissioning phase of the
  experiment; observations were made with three operational telescopes
  in late 2003 and with the complete 4 telescope array in January --
  February 2004 and October 2004 -- January 2005. } {Observations of
  the Crab nebula are discussed and used as an example to detail the
  flux and spectral analysis procedures of \hess.  The results are
  used to evaluate the systematic uncertainties in \hess\ flux
  measurements.}  {The Crab nebula data are analysed using standard
  \hess\ analysis procedures, which are described in detail. The flux
  and spectrum of \grs\ from the source are calculated on run-by-run
  and monthly time-scales, and a correction is applied for long-term
  variations in the detector sensitivity. Comparisons of the measured
  flux and spectrum over the observation period, along with the
  results from a number of different analysis procedures are used to
  estimate systematic uncertainties in the measurements.}  {The data,
  taken at a range of zenith angles between 45\dg\ and 65\dg, show a
  clear signal with over 7500 excess events. The energy spectrum is
  found to follow a power law with an exponential cutoff, with photon
  index $\Gamma = 2.39 \pm 0.03\stat$ and cutoff energy $E_{c} = (14.3
  \pm 2.1\stat) \textrm{TeV}$ between 440 GeV and 40 TeV. The observed
  integral flux above 1 TeV is $(2.26 \pm 0.08\stat) \times 10^{-11}
  \iflux$. The estimated systematic error on the flux measurement is
  estimated to be 20\%, while the estimated systematic error on the
  spectral slope is 0.1.}  {}

\keywords{Gamma rays: observations -- ISM: individual objects: Crab
  nebula -- ISM: plerions}

\maketitle

\section{Introduction}

The Crab supernova remnant (SNR) is an exceptionally well studied
object, with extensive observations of the system existing across the
entire accessible spectrum. At a distance of 2000 parsecs, with an age of
950 years, it is a prototypical centre-filled SNR, or plerion, as
defined by \citet{weiler80}. Within the supernova remnant lies the
Crab pulsar, with a rotational period of 33~ms and a spin-down
luminosity of $\textrm{L} = 5 \times\ 10^{38}\ \mathrm{erg} \ 
\mathrm{s}^{-1}$. This energy source powers a surrounding synchrotron
nebula, and polarization measurements exist from radio to hard X-ray
wavelengths \citep{wilson72}, indicating the non-thermal origin
of the radiation detected. The total energy available from the pulsar
to power the system is of the order of $10^{49}$~ergs.  This is
believed to be the power source for production of very high energy
(VHE) \grs.

The rotational energy of the pulsar is thought to be mostly carried
away by a relativistic wind of electrons and positrons. Interaction of
this wind with the surrounding medium causes a standing termination
shock wave \citep{rees74,kennel84}. Electron acceleration may be due
to a Fermi-type process \citep{achterberg01} or to driven reconnection of
the alternating magnetic field at this termination shock
\citep{coroniti90,michel94}.  The interaction of accelerated electrons
with ambient photon fields (in this case mostly synchrotron photons)
can produce VHE \grs\ via the inverse Compton process.

The Crab nebula was discovered at VHE energies in 1989
\citep{weekes89a} and emission has been confirmed by a number of other
experiments,
\citep{goret93,themistocle93,masterson99,aharonian00b,smith00,atkins03}.
Due the high flux from the source relative to other known TeV sources,
and its expected flux stability, it is conventionally used as a
standard reference source for VHE astronomy.  At the latitude of the
\hess\ experiment the Crab nebula culminates at 45\dg, so observations
of this source must always be made at large zenith angles; the greater
effective optical depth of the atmosphere increases the energy
threshold and affects the sensitivity of the detector. As the size of
the Crab nebula is small compared to the \hess\ point-spread function,
it may be treated as a point source for this analysis.

A detailed description of the \hess\ detector is given here, along with a
discussion of the principal sources of systematic error in the
atmospheric-Cherenkov technique. The \hess\ observations of the Crab
nebula are then detailed, and used as an example in a discussion of
the data calibration and analysis methods used in source
reconstruction and flux and spectral measurements. The stability of
the \gr\ flux and energy spectrum is measured using a number of
analysis methods, and a correction for variations in the long-term
optical efficiency of the detector is described. Using these results
the systematic uncertainties on flux measurements with \hess\ are
quantified. The sensitivity of the detector for source analysis is
also discussed. It should be noted that while the analysis presented
here is generally used to analyse targets observed by \hess, other
techniques are also used, e.g. \citet{lemoine06}.

\section {The \hess\ Experiment}

\hess\ is situated in the Khomas highlands of Namibia (23\dg\ 16'18''
South, 16\dg\ 30'00'' East), at an elevation of 1800 metres above sea
level. The four \hess\ telescopes are placed in a square formation
with a side length of 120 metres. This distance was optimised for
maximum sensitivity at the planned energy threshold of 100 GeV. 

\subsection{The detectors}

The \hess\ telescopes are of steel construction, with altitude/azimuth
mounts capable of precisely tracking any source from 0.0\dg\ to
89.9\dg\ in elevation, with a slew rate of 100\dg\ per minute
\citep{bolz04}. The dishes have a Davies-Cotton style hexagonal
arrangement \citep{davies57} with a flat-to-flat diameter of 13 m,
composed of 382 round mirrors, each 60 cm in diameter.  Thus the
effective mirror surface area is $107\ \textrm{m}^2$.  Further details
of the optical structure are given by \citet{bernlohr03}.  The mirrors
are remotely adjustable under computer control, and a fully automated
procedure is used, in conjunction with a CCD camera mounted in each
dish, for optimal alignment onto the focal plane of each telescope
camera, which is 15 m distant. Due to the rigidity of the dishes, this
alignment is stable over time scales in excess of one year. The
stability has been verified by regular monitoring of the optical point
spread function. Details of the mirror alignment system and the
optical point spread function are discussed by \citet{cornils03b}.

The \hess\ cameras each consist of a hexagonal array of 960 Photonis
XP2960 photo-multiplier tubes (PMTs). Each tube corresponds to an area
of 0.16\dg\ in diameter on the sky, and is equipped with Winston cones
to capture the light which would fall in between the PMTs, and also to
limit the field of view of each PMT in order to minimise background
light. The camera is of modular design, with the PMTs grouped in 60
\textit{drawers} of 16 tubes each \citep{vincent03}, which contain the
trigger and readout electronics for the tubes, as well as the high
voltage (HV) supply, control and monitoring electronics. The total
field of view of the detector is 5\dg\ in diameter.

The trigger system of the \hess\ array consists of three levels.
First, a single pixel trigger threshold is required, equivalent to 4
photo-electrons (p.e.) at the PMT cathode within an interval of 1.5
nanoseconds. Second, a coincidence of 3 triggered pixels is required
within a \textit{sector} - a square group of 64 pixels - in order to
trigger a camera. Each camera has 38 overlapping sectors.
Third, when the detector is operating in stereo mode, a coincidence of
two telescopes triggering within a window of (normally) 80 nanoseconds
is required.  Only cameras which have individually triggered are read
out in a stereo event. The stereo trigger system and the trigger
behavior of the \hess\ array is described by \citet{funk04}.

During the first and second level trigger formation, the individual
signals from each pixel are stored in two \emph{analogue ring sampler}
(ARS) circuits. A high gain and a low gain circuit are used to give
optimal signal dynamic range. The signals captured by each tube are
digitised in the drawer, before being collected by a central CPU in
the camera and sent to the central data acquisition system (DAQ)
by optical ethernet connection \citep{borgmeier03}.

The \hess\ experiment commenced observations in June 2002 with the
first telescope. The second telescope was installed in February 2003,
the hardware level stereo coincidence trigger was added in July 2003.
In the interim the two telescopes observed in parallel and an off-line
coincidence was used for stereo analysis. The third telescope was
installed in September. The full array was completed in December of
2003 and has been operational since then.

\subsection{Systematic uncertainties}
\label{sec:systematics}

The imaging atmospheric-Cherenkov technique depends on a form of
electromagnetic calorimetry to estimate the energy of observed
particles. In order to accurately measure the energy of the primary
particle which gives rise to an air shower, it is necessary to
understand the relationship between the particle energy and the signal
recorded in the cameras. Monte Carlo simulations of air showers in the
atmosphere are used to predict the light yield in the detector, and
thus the recorded signal, as a function of energy and shower position
relative to the observer.

There are three main contributions to uncertainties in the measured air shower
information, and thus the absolute flux calibration of the
detector:

\begin{enumerate}

\item The camera response. The single photo-electron response of each PMT
  varies strongly with the detector voltage and is measured using an
  LED system mounted in front of the camera. A second LED system
  mounted on each dish provides a uniform illumination across the
  camera and is used to correct for relative quantum efficiency
  variations of the PMTs, as well as in the reflectivity of the
  Winston cones in front of each PMT.  The complete calibration of the
  \hess\ telescope is described by \citet{aharonian04d}.

\item The optical response of the instrument, including the mirrors,
  Winston cones, shadowing by structural components of the system and
  the quantum efficiency of the photo-cathodes of the PMTs. This
  response can be measured by studying the Cherenkov light from single
  muons passing close to the telescope, assuming the camera response
  is well measured.  The use of muons in monitoring the telescope
  efficiencies is detailed by \citet{bolz04} and \citet{leroy04}. The
  optical response of the instrument degrades over a timescale of
  years as, for example, the mirror reflectivity decreases. This
  decrease in optical response, relative to that used in the Monte
  Carlo simulations, is taken into account in estimating the flux from
  a \gr\ source. This is described in detail in section
  \ref{sec:opcorr}.
  
\item The interactions of particles and light in the atmosphere. The
  atmosphere is the largest and least well understood component of a
  Cherenkov detector, being subject to variations in pressure,
  temperature and humidity. Two important effects of variability in
  the atmosphere are density profile variations, which affect directly
  the height of the shower maximum in the atmosphere and thus the
  intensity of the light seen at the telescope, and absorption of
  Cherenkov light in the atmosphere by clouds and dust, which leads
  directly to a reduction in the telescope trigger rates, and
  incorrect \gr\ energy reconstruction. These effects are discussed further
  by \citet{bernlohr00}.  Atmospheric monitoring devices are used to
  understand the local conditions under which the data has been
  recorded; such measurements are discussed by \citet{aye05}.
  Variations in the atmosphere can lead to rapid variability in the
  detector response, on a timescale of hours, so runs taken under
  variable conditions are rejected, as discussed in section
  \ref{sec:runsel}.
   
\end{enumerate}

\section{Observations}
\label{sec:obs}

\begin{table*}
\centering
\begin{tabular}{cccccccccc}
\hline
Data Set&Date    &N$_{\textrm{tels}}$&Z range      &$<\textrm{Z}>$ &Offset  &N$_{\textrm{runs}}$ &Obs. Time &Live-time &Mean System rate\\
        &        &                   &(\dg)        &(\dg)          &(\dg)   &                    &(hours)   &(hours)   &(Hz)\\
\hline                                                                           
I       &10/'03 - 1/'04 &3           &45-55        &46.6           &0.5-1.5 &12                  &5.3       &4.8       &179\\  
II      & 1/2004        &4           &45-65        &54.4           &0.5-1.5 &14                  &6.3       &5.7       &240\\
III     &10/'04 - 1/'05 &4           &45-55        &47.9           &0.5-1.5 &26                  &11.3      &10.6      &180\\ 
\hline
Total   &        &                   &45-65        &50.2           &0.5-1.5 &52                  &22.9      &21.1      &196\\
\hline
\end{tabular}
\caption{Details of the observations of the Crab nebula with \hess\ between October 2003 and January 2005. 
For the purposes of this study, the data has been divided into 3 subsets.}
\label{tab:obs}
\end{table*}

\subsection{The Crab data set}

The observations of the Crab nebula discussed in this paper are
summarised in Table \ref{tab:obs}. Data set I was taken between
October 2003 and January 2004 with three telescopes, and 5.3 hours of
data are included in this sample. These data range in zenith angles
from 45\dg\ to 55\dg. A further 6.3 hours (data set II) are included
from data taken in January 2004 with the complete array of four
telescopes, at a range of zenith angles from 45\dg\ to 65\dg.  Data
set III includes 11.3 hours, taken between October 2004 and January of
2005, also with 4 telescopes. All observations were taken using a
method (\emph{wobble} mode) whereby the source is alternately offset
by a small distance within the field of view, alternating between 28
minute runs in the positive and negative declination (or right
ascension) directions \citep{fomin94}.  This observation mode allows
the other side of the field of view, which does not contain the
source, to be used as a control region for estimation of the
background level.  The wobble offsets were varied from 0.5\dg\ to
1.5\dg\ for the data discussed here. The range of time over which the
observations have been taken allows us to study the long-term
stability of the \hess\ system.

Observations of the Crab were also taken in 2002 with a single
telescope, as discussed by \citet{masterson03}. These are not included
in this analysis, as the sensitivity is much lower than that in stereo
mode, and the systematic uncertainties of single-telescope
observations are greater than for stereo mode.

\subsection{Data quality selection}
\label{sec:runsel}

\begin{figure*}
  \centering $\begin{array}{c@{\hspace{0.1cm}}c}
  \multicolumn{1}{l}{\mbox{\bf }} & \multicolumn{1}{l}{\mbox{\bf }} \\
  [0cm]
  \resizebox{0.5\hsize}{!}{\includegraphics{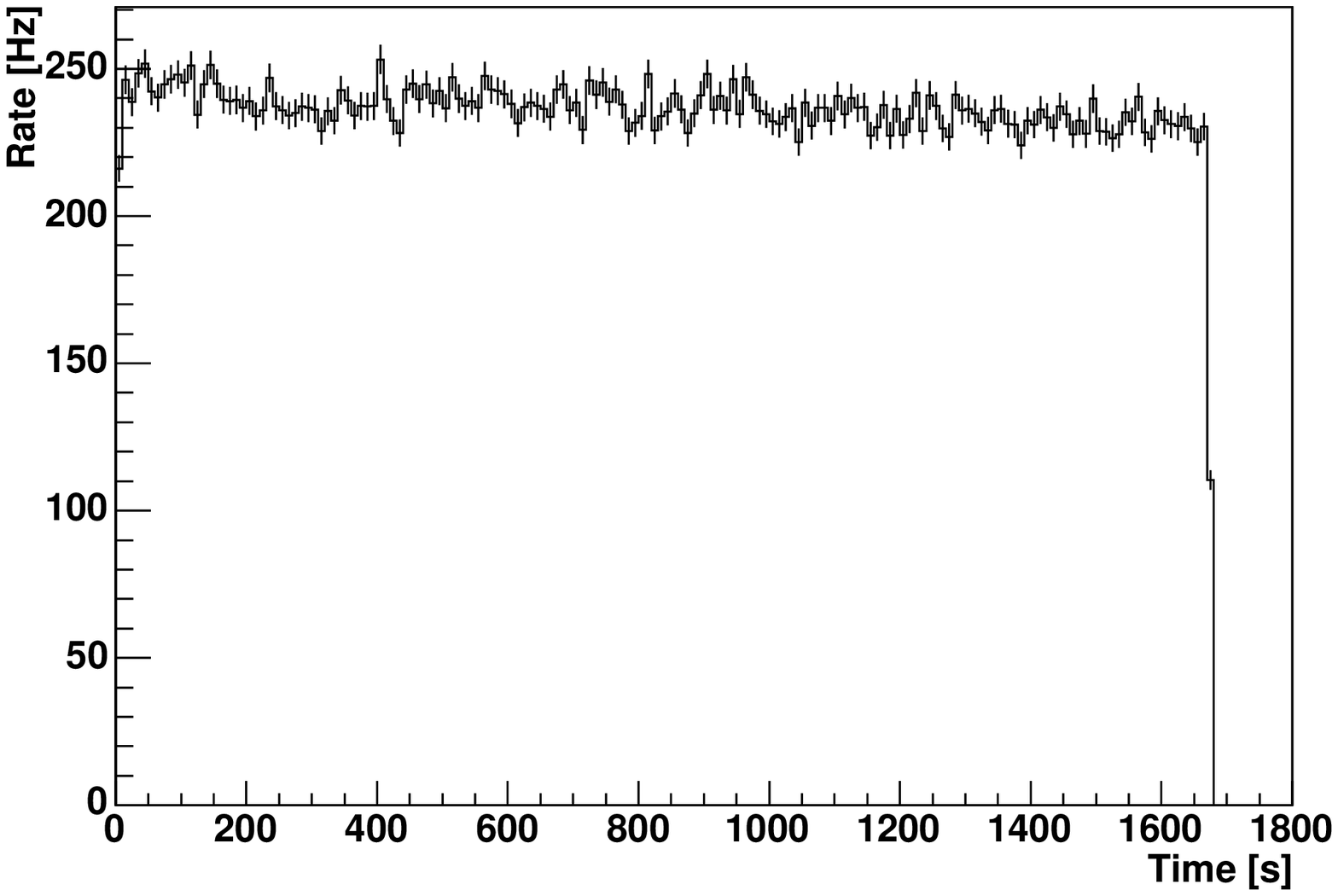}}&
  \resizebox{0.5\hsize}{!}{\includegraphics{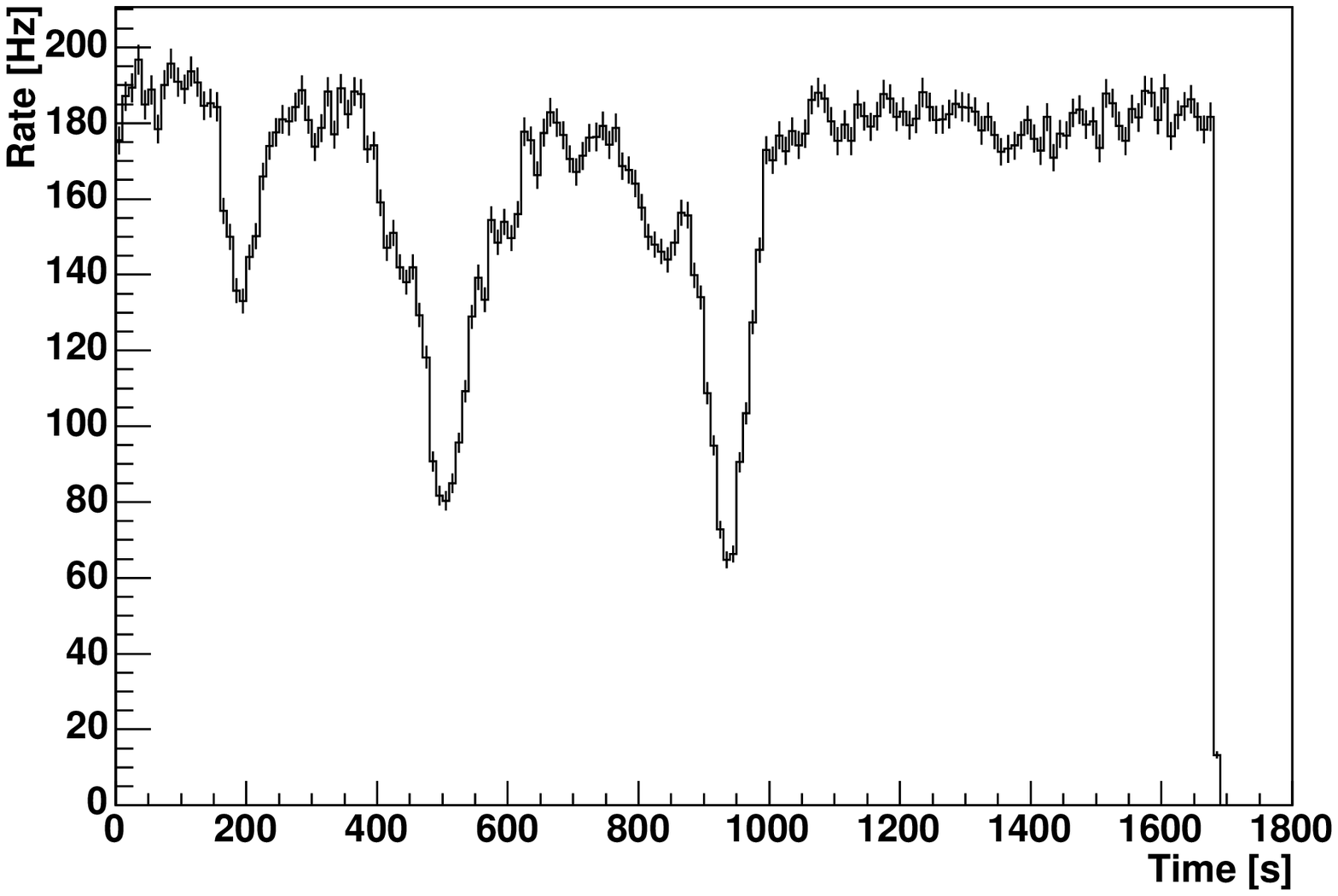}}
  \\ [0cm] \mbox{\bf (a)} & \mbox{\bf (b)} \end{array}$ 

\caption{The system rate vs time within a run for {\bf a)} a 4 telescope run
  passing the run selection, {\bf b)} a 3 telescope run failing the
  run selection. The run is rejected due to the instability in the
  rate caused by clouds passing through the field of view. The
  difference in absolute rate between the two runs is due to the
  differing zenith angles of the observations, as well as the number
  of telescopes active in each run.}
\label{fig:trigrate}
\end{figure*}

Systematic effects on the measured flux and energy spectrum (as
discussed in section \ref{sec:systematics}) may be ameliorated by
rejecting data recorded in non-optimal conditions.  Although the Crab
was observed for a total of nearly 45 hours, only 22.9 hours are
included in this analysis. The remaining observations have been
rejected as not meeting the run quality criteria. Some were made when
the sky conditions were less than optimal; the presence of clouds or
excessive dust in the atmosphere can lead to the absorption of
Cherenkov light and thus fluctuations in the system trigger
efficiency, causing systematic uncertainties in the measured \gr\ 
flux. Figure \ref{fig:trigrate} shows the trigger rate as a function
of time for two runs, one with a stable system trigger rate close to
the predicted level for this zenith angle, the other exhibiting
variability due to the presence of clouds.

Runs for which the mean trigger rate is less than 70\% of the
predicted value (as discussed by \citet{funk04}), or for which the rms
variation in the trigger rate is above 10\%, are rejected. As the Crab
is mainly visible during the rainy season in Namibia when weather
conditions are not optimal, the rejection rate for these data is not
representative of all \hess\ observations. The mean system rate is
240 Hz for the four telescope data and 180 Hz for the 3 telescope
data.

\begin{figure}
  \centering
  \includegraphics[width=0.5\textwidth]{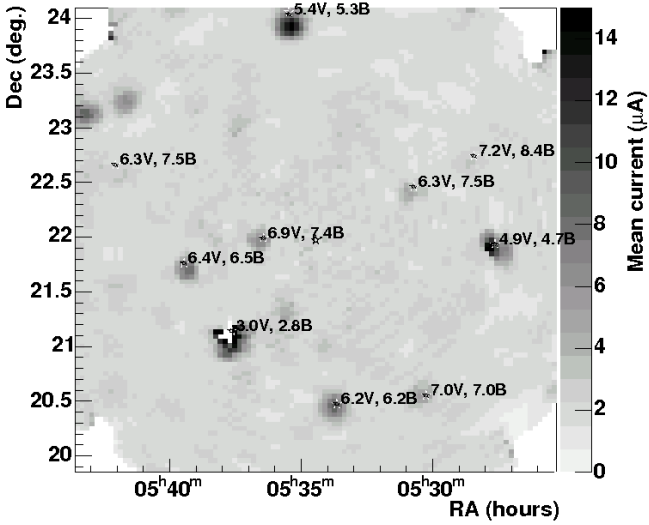}
  \caption{Sky map of DC pixel current (in units of $\mu$A, measured in a window
    of $16\,\mu$s) for the region surrounding the Crab nebula. The
    positions of stars (read from the Hipparcos catalogue -
    \citet{perryman97}) match well with the peaks in the map, which
    correspond to peaks in sky brightness. The B and V magnitudes of each
    star are also given for comparison. The brightest star, $\zeta$
    Tauri (B magnitude 2.8), has caused the camera to disable PMTs, thus
    there is no current measure for the region close to this star.}
  \label{fig:currentmap}
\end{figure}

Quality checks are also routinely carried out in order to reject runs
in which the array tracking system may not be functioning correctly,
leading to errors in the reconstructed position of the source, and
thus the flux. The nominal performance of the \hess\ tracking system
is discussed by \citet{hofmann03}. The tracking errors reported from
the DAQ are monitored and runs with rms deviations of more than 10
arcseconds in altitude or azimuth are rejected, in order to exclude
runs in which the tracking system malfunctioned.  However, no
observations on the Crab in this study (and in general very few runs)
have failed this test. As an independent check of the tracking quality
the DC PMT currents \citep{aharonian04d} are used to estimate the
amount of light impingent on each pixel during every run as a function
of time. A map is then made of the sky brightness in the field
of view of each telescope.  The positions of known stars are then
correlated with this map, giving a measure of the pointing position of
each telescope, independent of the tracking system and standard
pointing corrections.  A sample image for the region surrounding the
Crab nebula is shown in Figure \ref{fig:currentmap}.  Runs are
rejected if the pointing deviation is greater than 0.1\dg.  Again no
runs are found to have failed this test, which acts as an auxiliary
check of the tracking system.

The presence of bright stars in the field of view may trigger the
over-current protection of individual PMTs, causing them to be turned
off. Over-current due to bright star images is predicted and the
relevant PMTs are turned off in advance, for the duration of the star
transit through the PMT. Occasionally other bright, transient light
sources pass through the field of view of the telescopes. These are
normally meteorites, however lightning, airplanes and satellites may
also cause problems. The events cannot be predicted and thus trigger the
over-current protection, causing the PMT to be turned off for the
remainder of the run. The mean number of PMTs inactive during each
run for this and other hardware related reasons is monitored,
and telescopes with more than 10\% of the PMTs missing at any one time
are rejected from the analysis.  The PMT quality cut only rejects
occasional runs from each data set, and the mean number of pixels per
telescope disabled during the Crab nebula observations is 66, with an
rms of 7.

The observation time after run selection is corrected by taking into
account the dead-time of the system, when the trigger is not sensitive
to air showers. Estimation of the live-time (the time during an
observation in which the telescope system is sensitive to triggers
from the sky) is discussed by \cite{funk04}. The total live-time for
each data set is listed in Table \ref{tab:obs}. The systematic error
on this estimation is less than 1\%.

\subsection{Photoelectron to pixel amplitude calibration}

\begin{figure}
  \centering
  \includegraphics[width=0.5\textwidth]{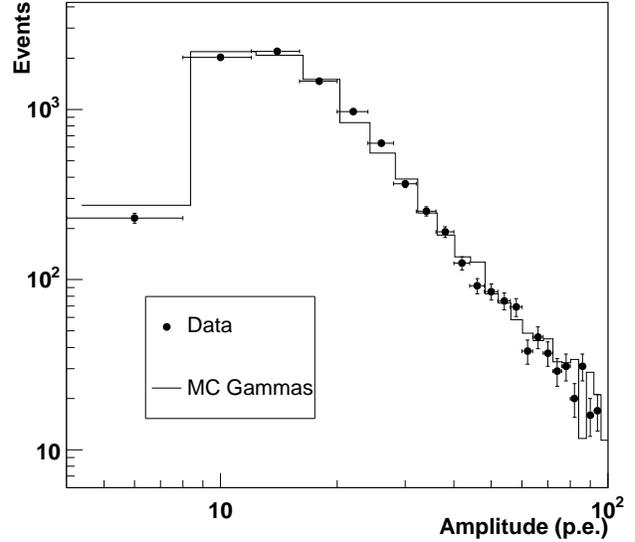}
  \caption{Distribution of the intensity of the 3rd highest pixel
    for 4 telescopes (points), compared with Monte Carlo simulations
    (solid lines). The overall flux of the simulated data has been
    adjusted to fit the real data.}
  \label{fig:pixel_int}
\end{figure}

All of the sources of uncertainty discussed in section
\ref{sec:systematics} affect the pixel intensities, and thus the
energy reconstruction of each shower. The trigger efficiency and
energy threshold are also affected, contributing to uncertainty in
absolute flux measurements. Part of this effect, namely the camera
response, is continuously calibrated using the LED system; the
accuracy of this calibration system can be evaluated by comparing the
reconstructed pixel intensities from Monte Carlo simulations of air
showers with similar measurements on real data. Figure
\ref{fig:pixel_int} compares the intensities in the pixel with the
third highest signal amplitude, as seen in real data for all
triggering events, with those from Monte Carlo simulations of protons
(which comprise a large majority of events detected) at the same
zenith angle.  As the trigger requirement is 3 pixels above a certain
amplitude, the amplitude distribution of the 3rd pixel shows the
effective pixel trigger level of each telescope. The peak in the
distribution is determined by the trigger threshold of the telescope
and the energy spectrum of the simulated and real events. It can be
seen that the Monte Carlo simulations reproduce well the data (\rchisq
= 15.5/23).

\subsection{Optical efficiency correction}
\label{sec:opcorr}

The optical efficiencies of the \hess\ telescopes change over time,
probable causes include degradation in the mirror reflectivity. This effect
happens on a timescale of years and can be monitored using images of
local muons in the field of view, for which the light yield can be
predicted, as discussed in section \ref{sec:systematics}.

\begin{figure}
  \centering
  \includegraphics[width=0.45\textwidth]{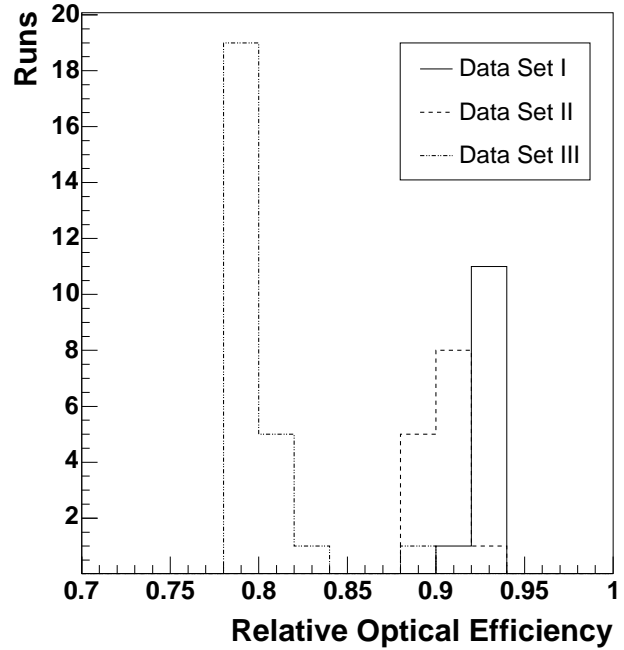}
  \caption{Distribution of the relative optical efficiency per run for
    data sets I (solid line), II (dashed line) and III (dotted line).
    It can be seen that the relative optical efficiency is
    significantly decreased in data set III relative to the other data
    sets.}
\label{fig:opcorr}
\end{figure}

The effect manifests itself as a reduction in the image intensity for
each event, compared to the intensity expected from Monte Carlo
simulations. This causes a shift in the absolute energy scale of the
detector, as events are reconstructed with energies which are too low.
This effect is corrected by incorporating a scaling factor into the
energy estimation for each event. The image intensity used in the
energy estimation is scaled by the ratio of the mean optical
efficiency (over the telescopes) for the run ($\textrm{Eff}_{run}$) to
the mean optical efficiency as derived from the Monte Carlo
simulations ($\textrm{Eff}_{mc}$). The corrected energy is then
derived from this scaled image intensity in the standard manner. The
distribution of the relative optical efficiency
($\frac{\textrm{Eff}_{run}}{\textrm{Eff}_{mc}}$) for the runs included
in this analysis is shown in Figure \ref{fig:opcorr}.  The application
of the optical efficiency correction in flux estimation is discussed
further in section \ref{sec:effarea}.

\section{Analysis}

\begin{figure}
  \centering
  \includegraphics[width=.48\textwidth]{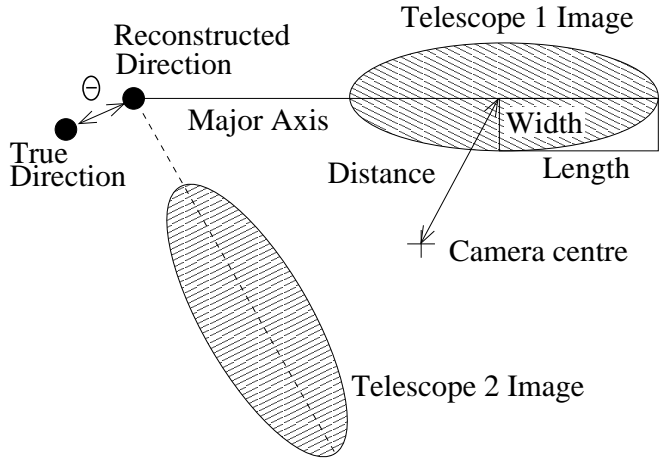}
  \caption{Definition of simple Hillas parameters, calculated for a
    \gr\ image, which may be approximated as an ellipse. Important
    parameters for this analysis are the width, length, distance. An
    image from a second telescope is superimposed to demonstrate the
    geometrical technique for source position reconstruction. The
    parameter $\theta$, which is the magnitude of the angular offset
    in shower direction reconstruction, is also shown.}
  \label{fig:par_dist}
\end{figure}

After a set of images of an air shower has been recorded, they are
processed to measure \textit{Hillas} parameters based on the second
moments of the image \citep{hillas85}. These parameters are then used
for event selection and reconstruction. A diagram illustrating the
parameter definitions is shown in Figure \ref{fig:par_dist}.

\subsection{Image cleaning and moment analysis}

The first step in the moment analysis procedure is image cleaning.
This is required in order to select only the pixels containing
Cherenkov light in an image. Other pixels, which contain mainly night
sky background (NSB) light are not used in the analysis. Images are
cleaned using a two-level filter, requiring pixels in the image to be
above a lower threshold of 5 p.e. and to have a neighbour above 10
p.e., and vice versa. Cleaning thresholds of 4 p.e. and 7 p.e. have
also been shown to work satisfactorily, but may be more sensitive to
uncertainties due to NSB light variations. This method selects
spatially correlated features in the image, which correspond to air
shower Cherenkov light.  This method tends to smooth out shower
fluctuations in a simple and repeatable manner.

After image cleaning, an image of a \gr\ shower approximates a narrow
elliptical shape, while images of background hadronic events are wider
and more uneven. The Hillas parameters are then calculated for each
cleaned image; these parameters are the basis for event selection.
The total amplitude of the image after cleaning is also calculated,
along with the mean position of the image in the camera, which
corresponds to the centroid of the ellipse.

\subsection{Stereo reconstruction}

The arrival direction of each event is reconstructed by tracing the
projected direction of the shower in the field of view (which
corresponds to the major axis of the image) to the point of origin of
the particle. For stereo observations it is possible to intersect the
major axes of the shower images in multiple cameras, providing a
simple geometric method of accurately measuring the shower direction;
more details, including methods to further improve the reconstruction
accuracy are given by \citet{hofmann99}, method I from that paper is
used here. Images are only used in the stereo reconstruction if they
pass the selection cuts on distance (to avoid camera-edge effects) and
image intensity. If less than two telescope images pass these cuts the
event is rejected.

\begin{figure}
  \centering
  \includegraphics[width=.48\textwidth,height=7cm]{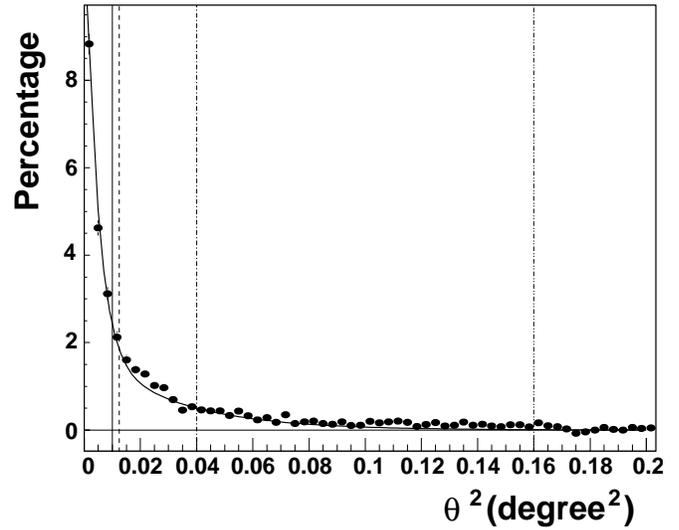}
  \caption{Distribution of excess events in \thsq\ for the complete Crab
    data set, after event selection and background subtraction. The
    Monte Carlo derived point-spread function described in equation 1
    is also shown, normalised to the excess distribution. The vertical
    lines denote the \thsq\ selection cuts listed in Table
    \ref{tab:selcuts}.}
  \label{fig:thsqonoff}
\end{figure}

Figure \ref{fig:thsqonoff} shows the excess distribution of \thsq\ for
data sets I-III, including events with two, three and four telescopes;
$\theta$ is defined in Figure \ref{fig:par_dist}, it is the angular
offset between the reconstructed shower direction and the true
direction of the Crab nebula. The distribution of reconstructed shower
directions is usually expressed in units of \thsq, as this ensures a
constant solid angle on the sky per bin. The value of the cut on
reconstructed shower direction is thus given in units of
$\mathrm{degrees}^{2}$ in Table \ref{tab:selcuts} for various sets of
selection cuts, and plotted in Figure \ref{fig:thsqonoff}. A strong
excess is seen close to zero, corresponding to events coming from the
direction of the Crab nebula. This distribution defines the accuracy
in the reconstructed arrival directions for \gr\ events from a point
source and is described by the \textit{point spread function} (PSF).
This function can be approximated by the sum of two, one-dimensional
Gaussian functions:

\begin{equation}
 PSF = \textrm{A} \left(\exp
  \left(\frac{-\theta^{2}}{2 \sigma_{1}^{2}} \right) + \textrm{A}_{rel}
  \exp \left( \frac{-\theta^{2}}{2 \sigma_{2}^{2}} \right) \right)
\end{equation}
 
This function is fitted to the \thsq\ distribution for simulated
Monte Carlo \grs. For simulations at 50\dg\ zenith angle the standard
deviation parameters $\sigma_{1}$ and $\sigma_{2}$ are 0.046\dg\ and
0.12\dg\ respectively.  The relative amplitude of the second Gaussian,
A$_{rel}$ is 0.15, while the absolute amplitude (A) is proportional to
the number of events in the fit.  The 68\% containment radius for
50\dg\ zenith angle is 0.12\dg, while that for 10\dg\ zenith angle is
0.10\dg. The point-spread function is shown in Figure
\ref{fig:thsqonoff} with the amplitude parameter (A) fit to the Crab
nebula data.  The \rchisq\ of this fit is $53/29$.

The position of the centre of the Cherenkov light pool, which
corresponds to the projected impact point of the original particle
track on the ground, can also be reconstructed by intersecting shower
axes, projected into the plane perpendicular to the system observing
direction. It is vital to reconstruct this position in order to
accurately measure the amount of light originally emitted by the
shower and thus the shower energy. The rms error on the reconstructed
impact parameter, which is the projected distance of the extrapolated
shower track to a telescope, for Monte Carlo simulations is less than
10 metres for events falling within 200 metres of the centre of the
array.

\subsection{Scaled parameter analysis}

The \textit{mean scaled width} method, similar to that used by the
HEGRA collaboration \citep{daum97}, is used to classify images as
either \gr\ like or hadron like, in order to reject non \gr\ 
background events. The main difference to the HEGRA method is in the
definition of the scaled parameter itself, in the HEGRA case this is
defined as $\mathrm{p}_{sc} = \mathrm{p}/ \langle \mathrm{p} \rangle$. A lookup table is used to
predict the mean width and length for a \gr\ as a function of the
amplitude of the shower image in the camera and impact parameter. Then
the value for a particular event (p) can be compared with the expected
value $\langle \mathrm{p} \rangle$ according to the formula:

\begin{equation}
  \mathrm{p}_{sc} = (\mathrm{p} - \langle \mathrm{p} \rangle)/\sigma_{\mathrm{p}}
\end{equation}

The mean value $\langle \mathrm{p} \rangle$ and the scatter $\sigma_{\mathrm{p}}$ for an
event vary with the image amplitude and impact distance, as well as
the zenith angle. Lookup tables are generated for 13 zenith angles (Z)
from 0\dg\ to 70\dg, based on Monte Carlo simulations. The
true impact parameter of the simulated shower is used in
filling the table.

When analysing real data, the reconstructed impact parameter is used
along with the image amplitude for each telescope image to find
$\langle \mathrm{p} \rangle$ and $\sigma_{\mathrm{p}}$ in the lookup
table. Linear interpolation (in $\cos (\mathrm{Z})$) between the two
nearest simulated values is then done to find the correct value for a
particular observation zenith angle. The \textit{mean reduced scaled
  width} (MRSW) and the \textit{mean reduced scaled length} (MRSL) are
then calculated by averaging over the telescope images passing the
image amplitude selection cut for each event: $\mathrm{MRSW} =
\left(\sum_{\mathrm{tel}}{\mathrm{p}_{sc}}\right) / \mathrm{N}_{\mathrm{tel}}$.

\begin{figure*}
\centering
  $\begin{array}{c@{\hspace{0.1cm}}c}
 
  \multicolumn{1}{l}{\mbox{\bf }} &
        \multicolumn{1}{l}{\mbox{\bf }} \\ [0cm]
  \resizebox{0.44\hsize}{!}{\includegraphics{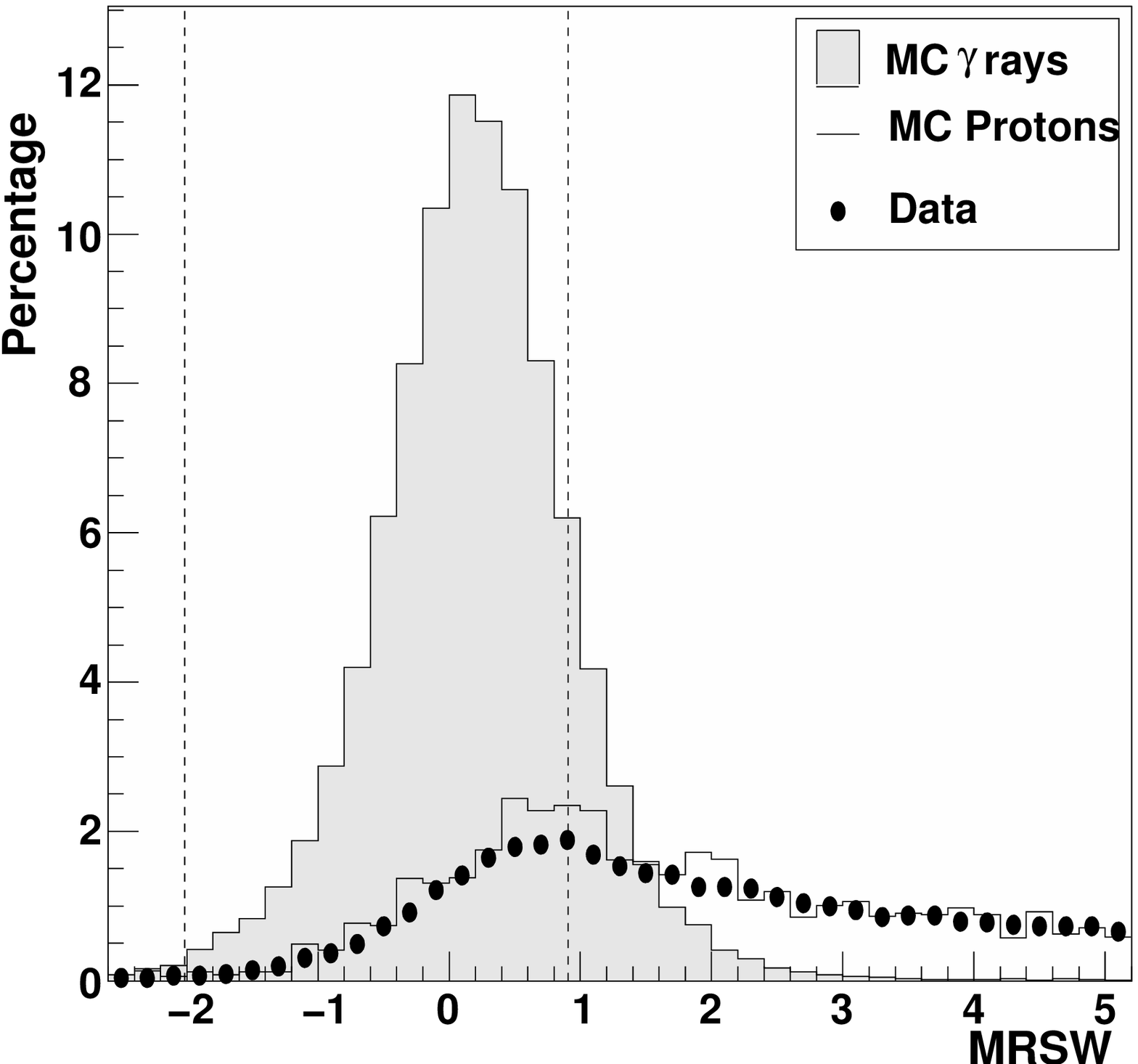}} &
  \resizebox{0.5\hsize}{!}{\includegraphics{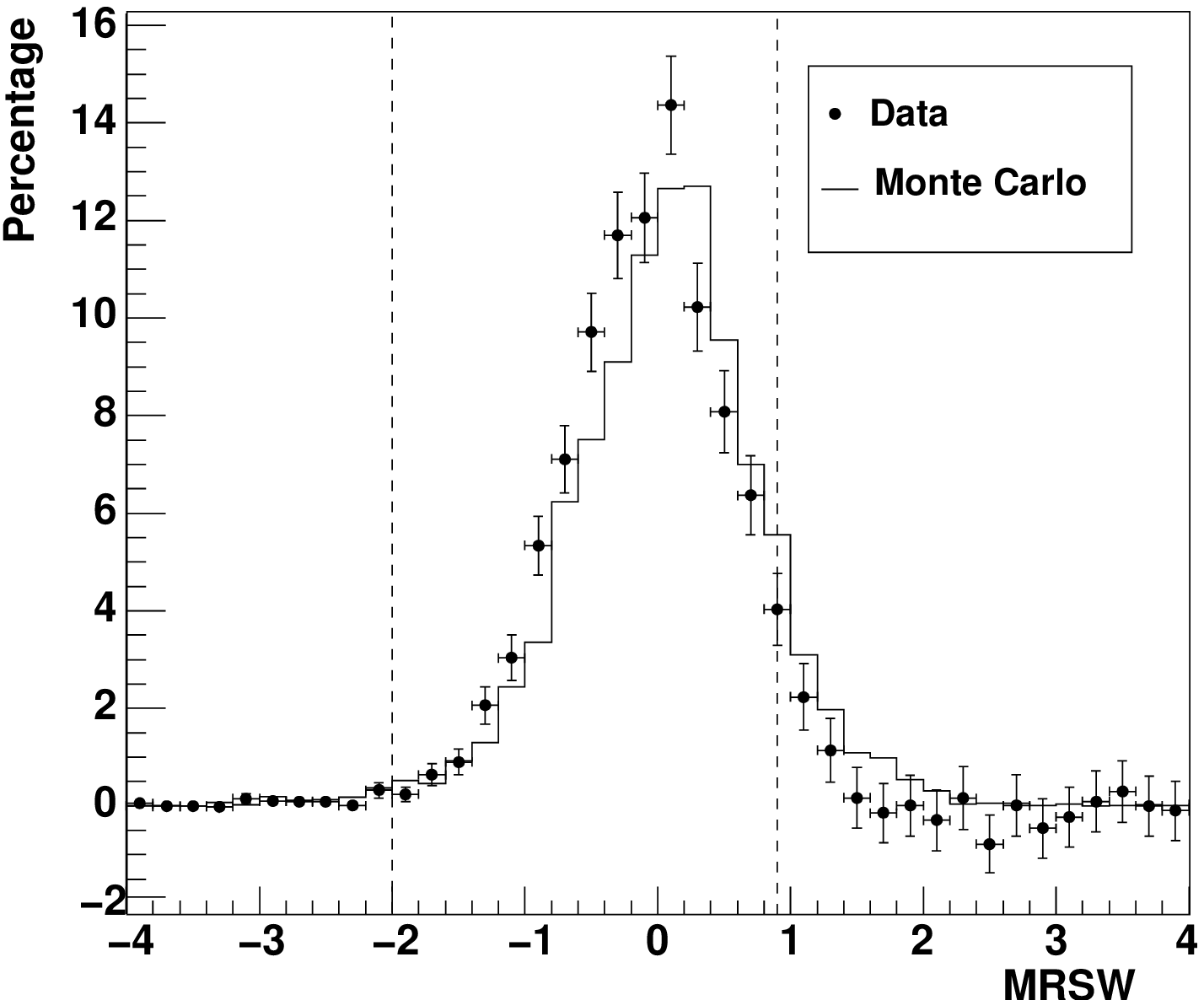}}\\ [0cm]
  \mbox{\bf (a)} & \mbox{\bf (b)}
  \end{array}$

  \caption{The distributions of mean reduced scaled width
    (MRSW) for Monte Carlo \gr\ simulations ($\Gamma$ = $2.59$) {\bf
      a)} with Monte Carlo proton simulations ($\Gamma$ = $2.70$) and
    actual \off\ data before selection cuts, {\bf b)} with real events
    from the direction of the Crab nebula (data set II) after
    selection cuts and background subtraction. All distributions are
    for zenith angle = 50\dg. The vertical lines denote the standard
    cuts described in Table \ref{tab:selcuts}.}
 \label{fig:params}
\end{figure*}

Figure \ref{fig:params} (a) shows a comparison between the MRSW
from Monte Carlo simulations of protons and \grs\ and from real data
at a zenith angle of 50\dg. It can be seen that the data (before
selection cuts) correspond well to Monte Carlo simulated protons, as
expected, while there is good separation between the data and Monte
Carlo simulated \grs, which are chosen to have a photon index
($\Gamma$) of 2.59, similar to that previously measured for the Crab
nebula \citep{aharonian00b}.

\subsection{Selection cuts}

Selection cuts on the mean scaled parameters, image intensity and
\thsq\ are simultaneously optimized to maximise the detection
significance ($\sigma$, as defined by \cite{li83}) for sources with
typical fluxes and energy spectra. The optimisation population
consists of a mixture of \gr\ simulations (selected to give the
desired flux and spectrum for optimisation) and real background data.
In the presence of background, the significance achieved for a given
source increases with the square root of the observation time;
instrument performance is therefore characterised by \sigrate. The
optimised cuts yield the maximum \sigrate\ for a source of that type.
It should be noted that the optimum selection cuts in any analysis
depend on the energy spectrum of the Monte Carlo simulations used in
the optimisation procedure, and it may be necessary to optimise
selection criteria separately for much harder or much softer energy
spectra. As a rule this is not done however in source searches, in
order to preserve the \textit{a priori} nature of the analysis. The
background data used in the optimisation is then not further used, in
order to avoid the possibility of optimising on background
fluctuations and compromising the statistical independence of the
results. The selection cuts are summarized in Table \ref{tab:selcuts}.
A number of alternative sets of cuts are presented here, which are
used for analysis of \hess\ sources. It will be shown that the
reconstructed flux and spectrum of the Crab are consistent for these
various selection criteria.

The standard set of selection cuts has been optimised to give the
maximum \sigrate\ for a flux 10\% of the Crab (\textit{standard}
cuts), with a similar spectrum.  The \textit{Hard} cuts are optimised
for a source with a flux 1\% of the Crab flux, and a $\Gamma$ of 2.0.
These cuts give a higher significance for weak, hard spectrum sources,
at the expense of energy threshold and cut efficiency. The hard cuts
are also useful as they reduce the systematic uncertainties in sky-map
reconstruction by reducing the numbers of background events, relative
to the signal. They also give a narrower PSF than the standard cuts,
as the higher intensity cut selects better reconstructed events. A set
of \textit{Loose} cuts have been also optimised to give the maximum
significance for a strong source, similar to the Crab, and a $\Gamma$
of 3.0. The lower intensity cut here reduces the energy threshold of
the analysis relative to the standard cut, and the fraction events
passing the cuts is higher. When conducting source searches, the
standard cuts are always used unless there is an \textit{a priori}
reason to expect a very hard or very soft spectrum from the source.

For analysis of large extended sources the cut on \thsq\ is usually
set to be larger than the extension of the source, so that effectively
all \grs\ from the source pass this cut.  In order to demonstrate the
effect of this strategy, a version of the standard cuts is described,
with the \thsq\ cut set to a much larger value. These are referred to
as \emph{extended} cuts for the purposes of this paper. It should be
emphasised that only the standard selection cuts are used in searches
for point sources, extended source searches are carried out using an
\emph{a priori} \thsq\ cut suited to the source size, and trials are
taken into account when testing multiple source extensions.

Figure \ref{fig:params} shows the distributions of MRSW, after
standard selection cuts and background subtraction (see following
section), for real data and simulations (with the same mean zenith
angle). The cut on MRSW is not applied for this plot. The standard
MRSW selection cuts are indicated, it can be seen that the cuts select
\gr-like events.  The small shift between the data and Monte Carlo
simulations seen in this plot is due to differences in the optical
efficiency; simulations with reduced efficiency (as in Figure
\ref{fig:opcorr}) agree well with the data. It can be seen that this
shift has a negligible effect on the efficiency of the scaled
parameter cuts.

\begin{table*}
\centering
\begin{tabular}{cccccccc}
\hline
Configuration&MRSL&MRSL&MRSW&MRSW &$\theta_{cut}^{2}$ &Image Amp. &Distance\\
             &Min. &Max. &Min. &Max. &Max.          &Min.        &Max.     \\
             &    &    &    &    &(degrees$^{2}$) &(p.e.)     &(\dg)  \\
\hline
Standard     &-2.0&2.0 &-2.0&0.9 &0.0125       &80        &2.0\\
Hard         &-2.0&2.0 &-2.0&0.7 &0.01         &200       &2.0\\
Loose        &-2.0&2.0 &-2.0&1.2 &0.04         &40        &2.0\\
Extended     &-2.0&2.0 &-2.0&0.9 &0.16         &80        &2.0\\
\hline
\end{tabular}
\caption{Selection cuts optimised for various purposes, as described
in the text. Cuts are applied on MRSW and MRSL, as well as on the distance
($\theta$) from the reconstructed shower position to the source. A
minimum of two telescopes passing the per-telescope cuts, on image
amplitude and distance from the centre of the field of view, are also
required. Standard cuts, as well as \emph{hard}, \emph{loose} and
\emph{extended} cuts, as described in the text, are listed.}
\label{tab:selcuts}
\end{table*}

\section{Signal extraction and background estimation}

\begin{figure}
  \centering
  \includegraphics[width=0.5\textwidth]{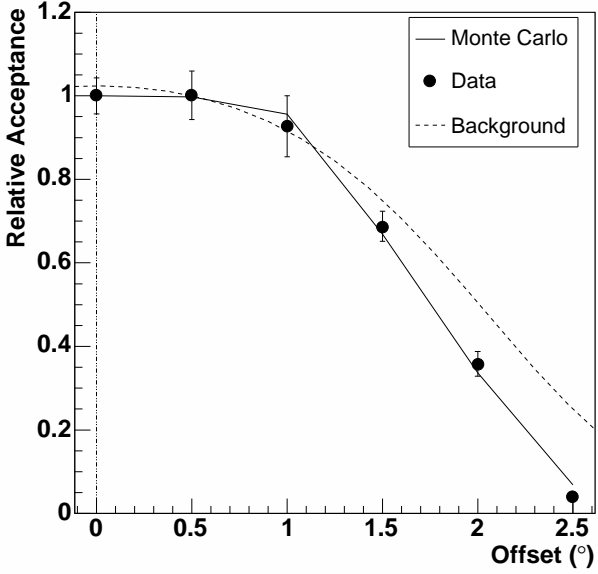}

  \caption{ A comparison of the predicted relative \gr\ rates (from MC simulations) and
    those measured from data (from 4-telescope test observations on
    the Crab nebula), as a function of off-axis angle in the field of
    view. Also plotted is the relative radial acceptance for
    background events passing selection cuts. This agrees well with
    the relative \gr\ acceptance out to 1.5\dg.}
\label{fig:rate_offset}
\end{figure}

When estimating the flux of \grs\ from a particular direction in the
sky it is necessary to estimate the background level, due to non \gr\ 
events with directions reconstructed close to the source direction.
The significance of the excess after background subtraction is then
determined using the likelihood ratio method described by
\citet{li83}. For the purpose of background estimation the
distribution of background events is usually assumed to be azimuthally
symmetric within the camera field of view. However, zenith angle
dependent effects or variations in the NSB level across the field of
view may introduce non-radial variations in the background level. A
radial profile of the relative rate of background events passing shape
cuts (the \emph{background acceptance}) in the field of view is shown
in Figure \ref{fig:rate_offset} (dashed line). For comparison, a
number of test observations (duration 30 minutes each) were made at a
range of offsets from 0\dg\ to 2.5\dg\ on the Crab nebula. It can be
seen that the relative rate of excess \gr\ events passing cuts
(points) follow the background acceptance closely out to 1.5\dg\ 
offset in the camera. The Monte Carlo predicted \gr\ rate for this
zenith angle is also shown (solid line), this is described in section
\ref{sec:effarea}.

When estimating the background, first the reconstructed shower
direction for each $\gamma$-like event (i.e. an event that passed the
shape cuts) is filled in a two dimensional histogram (so-called raw
\emph{sky-map}).  The \on\ signal for a given point in the sky is
determined by selecting events within a circle around that point with
radius $\theta_{cut}$. Two techniques are used to derive estimates of
the background level within this region of the field of view, and are
described below.

\subsection{Reflected background model}

The simplest background estimation technique uses the signal at a
single position in the raw sky-map, offset in the opposite direction
relative to the centre of the field of view, to estimate the
background. This technique is used in the standard \textit{wobble}
observation mode (described in section \ref{sec:obs}). However, it
suffers from relatively low statistics in the measurement of the
background level due to the choice of a single reflected background
position, as well as possible systematic effects caused by local
inhomogeneities at the background position.

\begin{figure}
  \centering
  \includegraphics[width=0.45\textwidth]{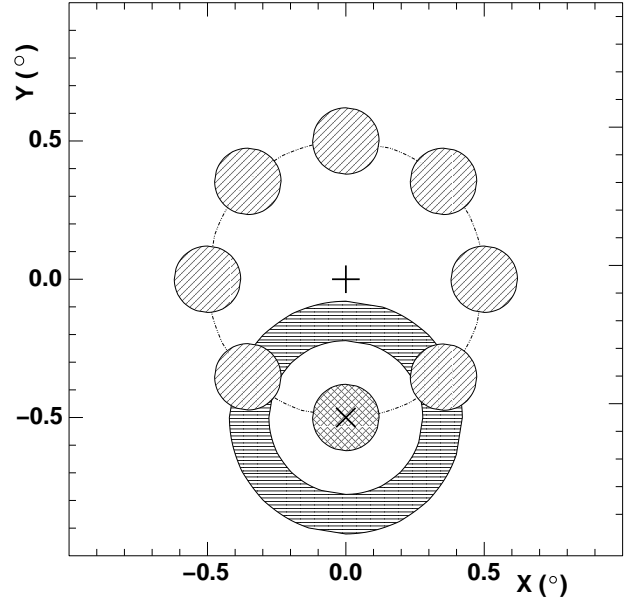}
  \caption{Schematic illustrating the background regions described in
    the text. The observation position of the telescopes is marked by
    a cross, while the target position is marked by an X. The \on\ 
    region surrounding the target position is marked by a cross
    hatched circle. The ring-background region is filled by horizontal
    lines, while the reflected-background regions, at constant offset
    from the observation position, are filled by diagonal lines. The
    two background regions have equal area in this case.}
\label{fig:backpos}
\end{figure}

The generalised reflected background technique, which is also suitable
for wobble mode, uses a number of background regions equidistant
from the observation position, as illustrated in Figure
\ref{fig:backpos}. The combined events from these positions are used
to estimate the background at the \on\ position, scaled by the
relative area of the \on\ and \off\ regions. In the case of a larger
\on\ integration region the number of background regions is reduced to
eliminate overlapping.  The normalisation, $\alpha$, is the ratio of
the solid angles of the \on\ and \off\ regions. As the \off\ positions
are the same distance from the centre of the field of view as the \on\ 
position, correction for the relative radial background acceptance of
the detector is not required.  However, this method cannot be used for
positions closer to the centre of the field of view than the radius of
the \on\ region, as the background positions would overlap with the
source position. As all of the data described here is taken in
\emph{wobble} mode, this method is used in this analysis for flux and
spectral measurements.

The reflected-background method may also be susceptible to systematic
effects caused by non-radial variations in the acceptance, especially
for large offset positions in the field of view. Non-radial effects
are strongest for data with only two telescopes, where the trigger
efficiency can vary with the azimuth direction of the shower impact
point on the ground relative to the telescope positions. This could be
corrected by applying a small non-radial correction term across the
field of view. This correction has not been applied in the analysis
described here. However, in order to minimise systematic effects due
to non-radial acceptance variations, the direction of the wobble
offset is generally alternated run by run on either side of the target
position.

\subsection{Ring-background model}

\begin{figure*}
  \centering $\begin{array}{c@{\hspace{0.1cm}}c}
    \multicolumn{1}{l}{\mbox{\bf }} & \multicolumn{1}{l}{\mbox{\bf }} \\
    [0cm]
    \resizebox{0.5\hsize}{!}{\includegraphics{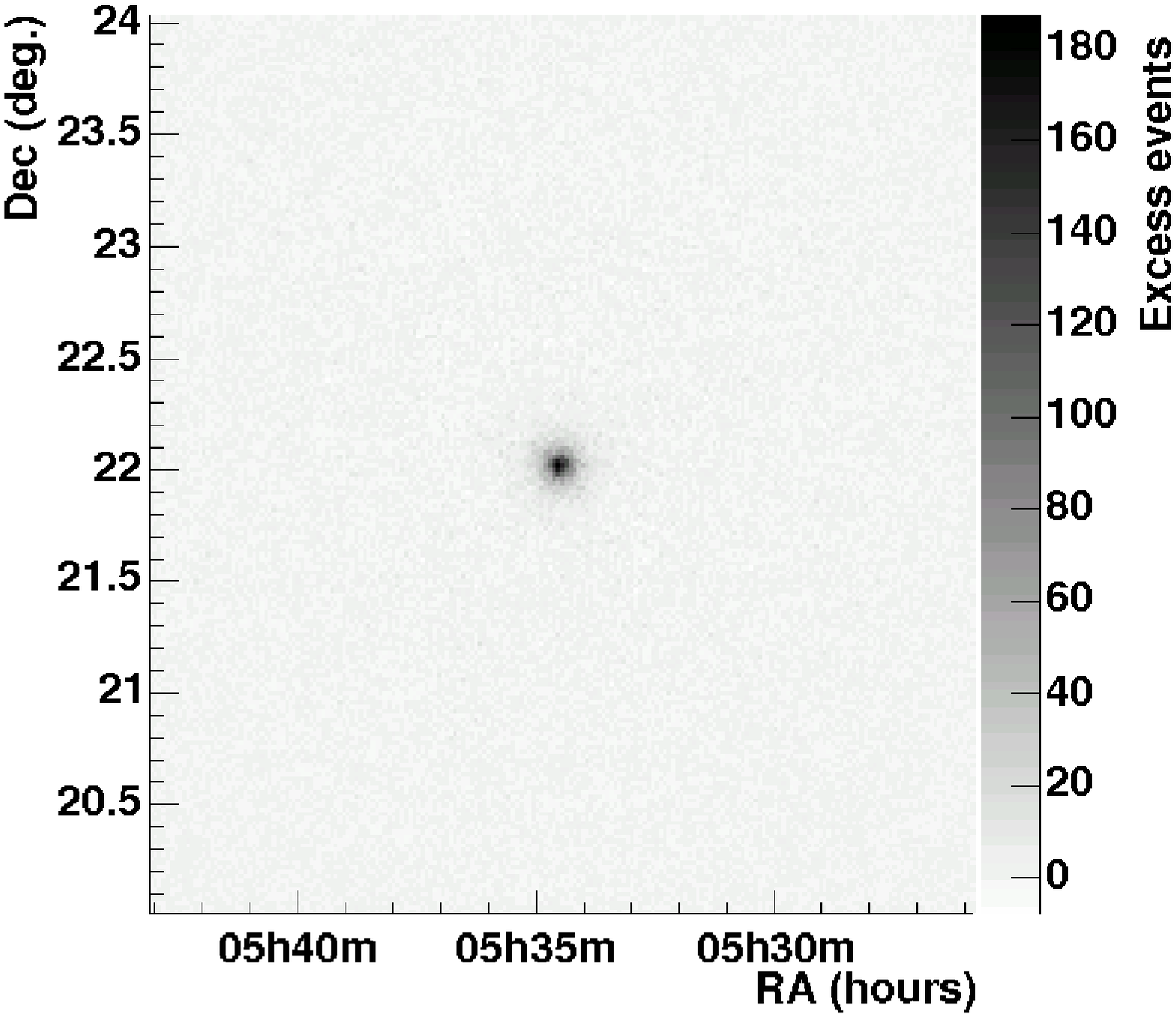}}
    &
    \resizebox{0.5\hsize}{!}{\includegraphics{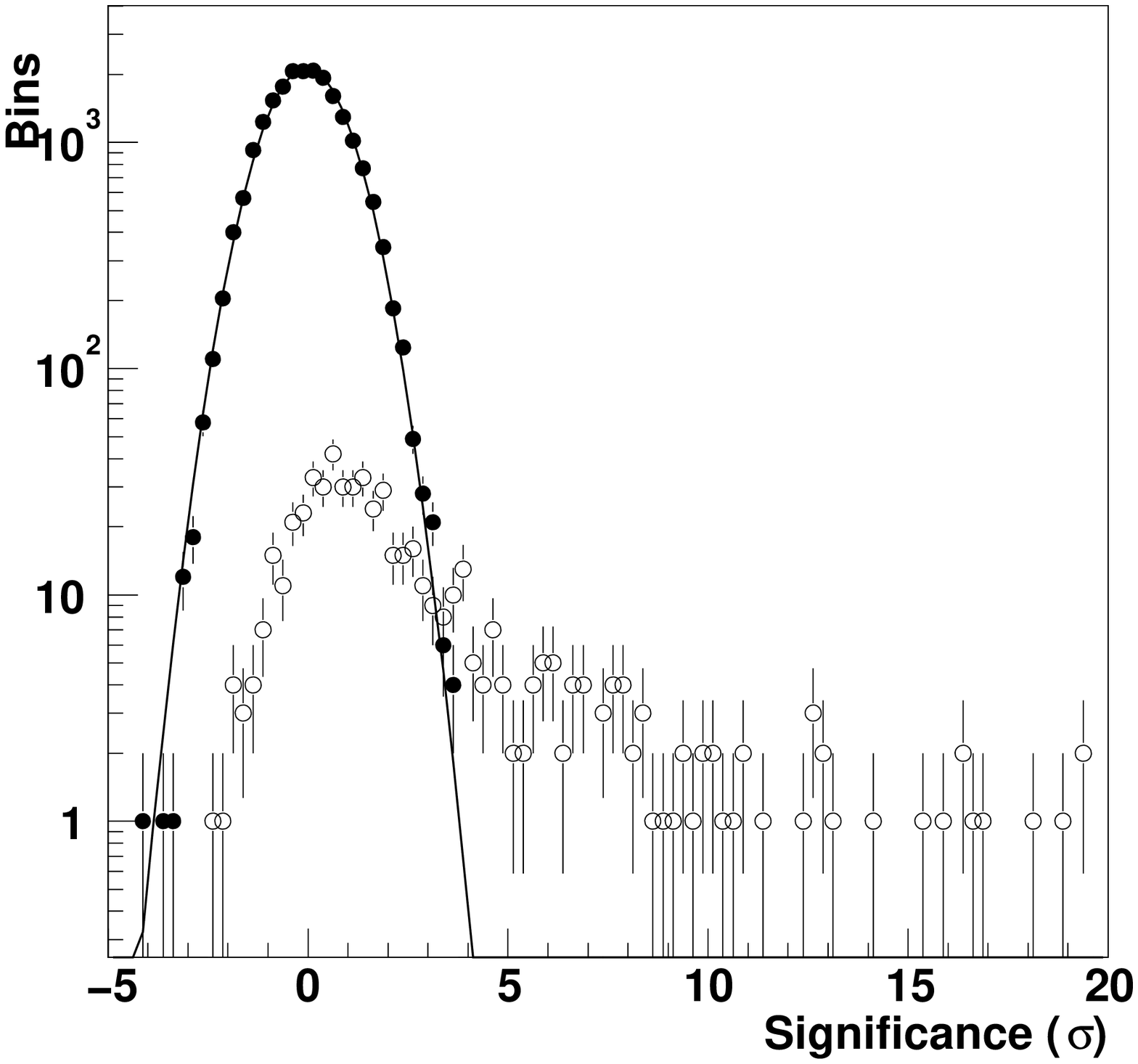}}
    \\ [0cm] \mbox{\bf (a)} & \mbox{\bf (b)} \end{array}$ \caption{
    {\bf a)} Uncorrelated 2-d plot of \gr\ excess from the Crab nebula, the
    reconstructed sky-map in RADec J2000 coordinates around the source
    position is shown.  {\bf(b)} The distribution of significance per
    bin in an uncorrelated significance sky map for the same data. The
    open circles denote the distribution for points in the map within
    0.4\dg\ of the Crab position, while the filled circles represent points
    further from the source. The Gaussian function, fitted to the
    second distribution, has a mean of $0.04 \pm 0.006\stat$ and a standard
    deviation of $0.98 \pm 0.004\stat$. A mean of 0.0 and standard
    deviation of 1.0 is expected for an unbiased significance
    distribution of the background.}
  \label{fig:2don}
\end{figure*}

The ring-background technique determines the background for each
position in the field of view using the background rate contained in a
ring around that position \citep{puehlhofer03}. The internal and
external radii of the ring are here chosen such that the ratio of the
areas of the \off\ to \on\ regions is close to 7, which makes for a
convenient compromise between area within the ring and distance from
the \on\ position. The inner ring radius is chosen to be significantly
larger than the on region, in order to avoid signal leakage into the
\off\ region. The normalisation ($\alpha$) is given by the area ratio
modified by a weight factor to account for the radial background
acceptance in the camera. The ring around the \on\ position is
illustrated in Figure \ref{fig:backpos}. When estimating the
background for a test position close to a known source like the Crab
nebula, the source position is cut out of the background ring in order
to avoid signal pollution in the \off\ region for the test position.
This method has the advantage of allowing background estimation for
all positions in the field of view. However, since the number of
events at positions surrounding the source is used to estimate the
background in the direction of the source, it is most suitable for
sources with an small angular extent relative to the field of view of
the detector. Figure \ref{fig:2don}(a) shows an excess map of the sky
in the vicinity of the Crab nebula, after background subtraction.
Figure \ref{fig:2don}(b) shows the distribution of significance of the
excesses in each bin in the sky map. It can be seen that the
significance is distributed normally in the off source regions of the
map (filled circles), while the region close to the Crab nebula (open
circles) shows a significant excess.

The ring-background method is less suitable for spectral analysis of
sources than the reflected method as the background acceptance may not
be constant as a function of energy, thus the background level may not
be correctly estimated for the entire energy range of the spectral
analysis. Thus this method is not used for the main spectral and flux
analysis in this paper, however results are derived using this method
for comparison purposes.

\section{Energy reconstruction and effective areas}

\subsection{Energy reconstruction}

The energy of the primary particle of a \gr\ shower is estimated for
each telescope as a function of the image amplitude and impact
parameter using a lookup table. The lookup table contains the mean
energy for Monte Carlo \gr\ simulations as a function of total image
amplitude and the simulated true impact parameter. As for the scaled parameters,
the lookup tables are created for a number of zenith angles and the
resulting energy is estimated by linear interpolation in $\cos (\mathrm{Z})$, and
averaged over the triggered telescopes for each event.  Events with
relative error in the reconstructed impact parameter greater than 25\%
are not used in the lookup table creation, in order to reject poorly
reconstructed events, which may bias the lookup table. Events with a
distance greater than 2\dg\ from the centre of the field of view are
also rejected.

In the case where no estimate is available in the lookup table for a
particular event, due to a lack of Monte Carlo statistics at that
combination of impact distance and total amplitude, an alternative
lookup is used with coarser impact distance binning. This occurs on
average for 0.3\% of events.  In the case where the optical correction
is applied, the energy is estimated using the corrected image
intensity, as described in section \ref{sec:opcorr}. The energy
estimated for each telescope is averaged to give the mean energy for
an event: $\mathrm{E}_{est} =
\left(\sum_{\mathrm{tel}}{\mathrm{E}_{\mathrm{tel}}}\right) /
\mathrm{N}_{\mathrm{tel}}$.

\subsection{Energy resolution}

\begin{figure}
  \centering
  \includegraphics[width=0.5\textwidth]{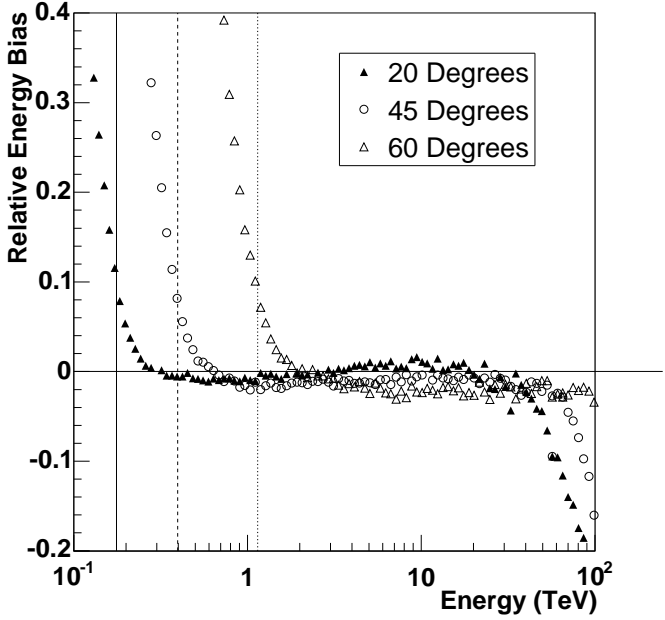}
  \caption{Relative bias ($(\mathrm{E}_{reco}-\mathrm{E}_{true})/\mathrm{E}_{true}$) in estimation
    of energy as a function of energy for three zenith angles. The
    vertical lines represent the safe energy thresholds for spectral
    analysis at each zenith angle.}
\label{fig:bias}
\end{figure}

The error in the reconstructed energy ($\Delta_{\mathrm{E}}$) for a
particular simulated \gr\ event with true energy $\mathrm{E}_{true}$
and reconstructed energy $\mathrm{E}_{reco}$ is defined as
$\Delta_{\mathrm{E}} =
(\mathrm{E}_{reco}-\mathrm{E}_{true})/\mathrm{E}_{true}$. The mean
value of $\Delta_{\mathrm{E}}$ is shown as a function of
$\mathrm{E}_{true}$ in Figure \ref{fig:bias}. For energies close to
the threshold, there is a bias due to a selection effect, whereby
events with energies reconstructed with too high a value are selected.
In order to make an accurate energy spectrum it is necessary to define
the useful energy range, so as to avoid the region of large energy
bias. First the lowest energy bin in the bias histogram with a bias of
less than 10\% is found. The useful lower energy threshold is the
maximum energy of this bin plus 10\%. This safe energy threshold is
also indicated in Figure \ref{fig:bias} for each zenith angle. This
energy threshold for the analysis is increased to take account of the
shift in the energy scale when the optical efficiency correction is
applied. It can be seen that the energy bias above the safe threshold
does not depend on the zenith angle, up to energies in excess of 60
TeV.

\begin{figure}
  \centering
  \includegraphics[width=0.5\textwidth]{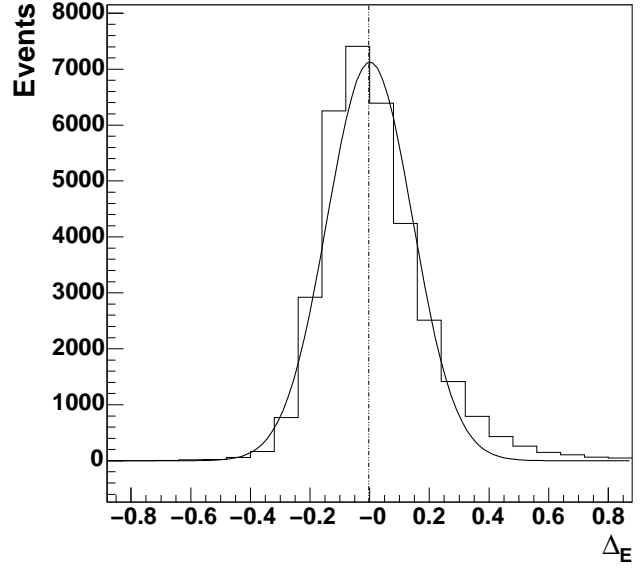}
  \caption{The distribution of the relative error in the reconstructed energy
  per event for Monte Carlo simulated \grs\ with a power law energy
  distribution (above 440 GeV) with $\Gamma = 2.6$ at 50\dg\ 
  zenith angle. The root mean square (rms) width of this distribution
  is 16\%. The width of the fitted Gaussian distribution is 14\%.}
\label{fig:energyres}
\end{figure}

The distribution of $\Delta_{\mathrm{E}}$, for \grs\ simulated with a power law
spectrum with $\Gamma = 2.6$, at 50\dg\ zenith angle, is shown in
Figure \ref{fig:energyres} for the standard analysis. The energy
resolution for a particular energy range is defined as the width of
this distribution. Events in this plot are selected above the safe
threshold of 0.44 TeV in order to avoid the effect of the energy bias.
The energy resolution defines the optimum binning for spectral
reconstruction, as well as defining the minimum energy width of any
resolvable spectral structure.  It is possible to improve the energy
resolution slightly by selecting only those events with higher
telescope multiplicities and smaller impact parameters, at the expense
of reduced event statistics.

\begin{figure*}
  \centering $\begin{array}{c@{\hspace{0.1cm}}c}
    \multicolumn{1}{l}{\mbox{\bf }} &
    \multicolumn{1}{l}{\mbox{\bf }} \\ [0cm]

    \resizebox{0.485\hsize}{!}{\includegraphics{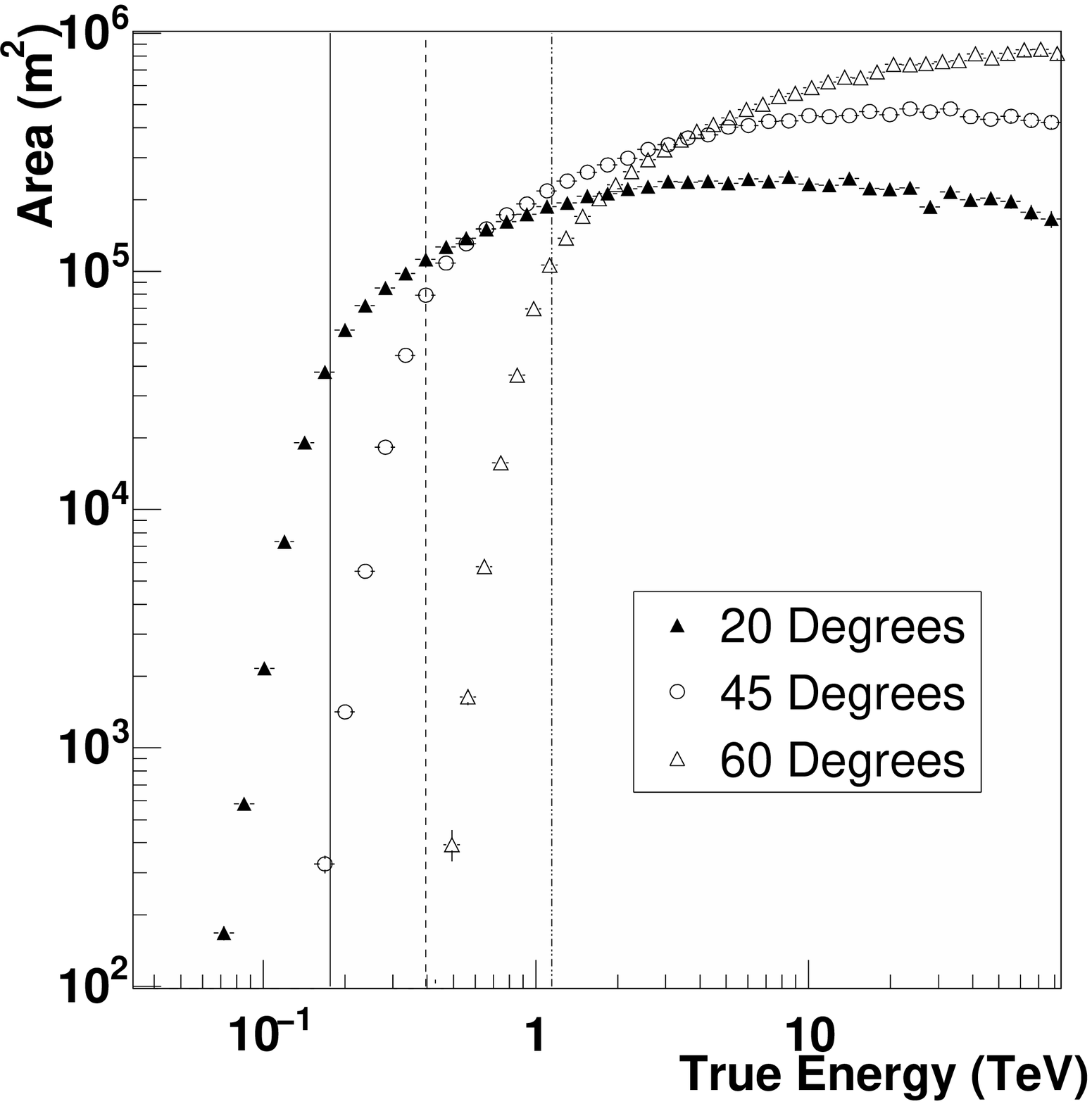}}
    &
    \resizebox{0.5\hsize}{!}{\includegraphics{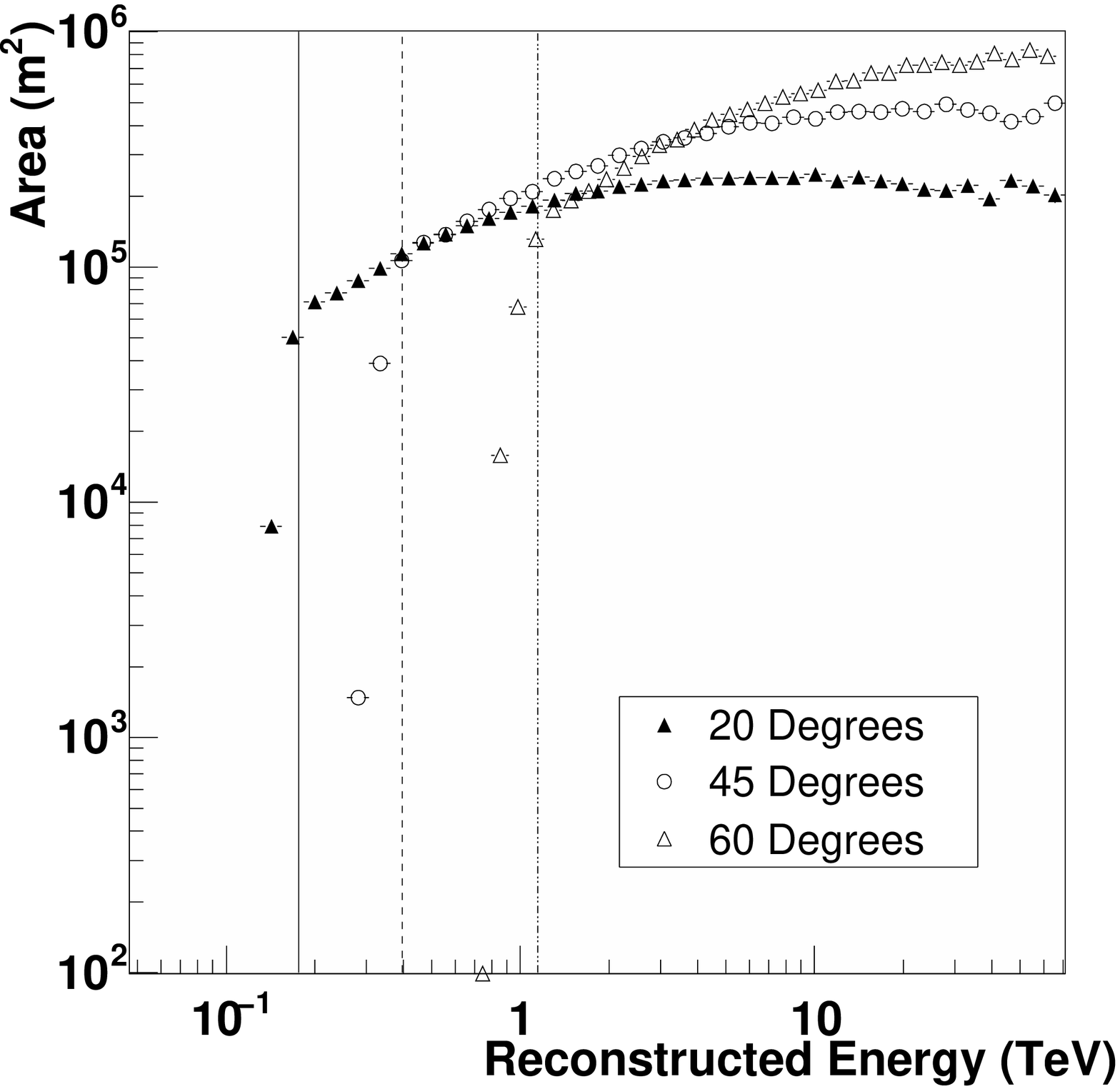}} \\ [0cm]
    \mbox{\bf (a)} & \mbox{\bf (b)}
  \end{array}$
 \caption{The effective collecting area of the full \hess\ array
 versus energy {\bf a)} as a function of true Monte Carlo energy and
 {\bf b)} as a function of reconstructed energy for observations at
 zenith angles of $20\dg$, $45\dg$, and $65\dg$. The vertical lines
 denote the safe energy thresholds for each zenith angle, increasing in
 zenith angle from left to right.}
\label{fig:effareas}
\end{figure*}

\subsection{Effective areas}
\label{sec:effarea}

The \gr\ flux from a source is estimated from the number of excess
events passing the selection cuts for a particular data set using the
effective area of the instrument. The effective area is a function of
the zenith angle and offset of the source from the pointing direction,
the energy of the event and the particular selection cuts used.  The
effective area is modeled from Monte Carlo simulations by counting the
fraction of simulated events which trigger the detector and pass the
selection cuts. This effective area has been estimated in two ways: as
a function of the Monte Carlo energy of the simulated events (\atrue)
and as a function of the reconstructed energy (\areco). While \atrue\ 
does not depend on the energy spectrum of the simulated \grs, the
finite energy resolution makes \areco\ sensitive to this; the
effective areas are usually estimated assuming a power law
distribution of $\Gamma = 2.0$.  However, when estimating a flux using
events binned in reconstructed energy, it is correct to use \areco\ to
estimate the effective area for each bin, when estimating the
integrated effective area over the whole energy range one may use \atrue. In
order to avoid a bias in the spectral reconstruction when using
\atrue, it is necessary for the energy spectrum of the simulations to
match that of the data, this is discussed further below. The effective
areas as a function of true energy and reconstructed energy (for the
standard selection cuts) are shown in Figure \ref{fig:effareas} for
three zenith angles. As the effective area of the telescope system
depends strongly on the zenith angle of the observations, it is
determined for a range of angles and the value for a particular energy
and zenith angle is determined by linear interpolation in $\log(E)$
and $\cos(\mathrm{Z})$ \citep{aharonian99a}.

\begin{table*}
\centering
  \begin{tabular}{c|cc|ccc|ccc|ccc|}
\hline
         &Pre-cut    &           &           &Standard cuts &  &            & Loose cuts  &      &         &Hard cuts  &       \\
Z        &Threshold  &rate       &Threshold  &rate      &      &Threshold    &rate        &       &Threshold  &rate    &       \\
(\dg)    &(TeV)      &(\pmin)    &(TeV)      &(\pmin)   &(\%)  &(TeV)        &(\pmin)     &(\%)   &(TeV)      &(\pmin) &(\%)   \\
\hline                                                                                                                          
0        &0.09       &56.5       &0.16       &19.8      &35    &0.13        &38.0         &67    &0.28        &7.8     &14     \\
20       &0.11       &52.8       &0.18       &19.4      &37    &0.15        &37.1         &70    &0.33        &7.4     &14     \\
30       &0.14       &45.7       &0.22       &16.6      &36    &0.19        &31.7         &69    &0.42        &6.3     &14     \\
40       &0.19       &35.1       &0.31       &12.1      &35    &0.26        &23.3         &66    &0.61        &4.6     &13     \\
45       &0.25       &29.2       &0.40       &9.44      &32    &0.33        &18.4         &63    &0.77        &3.6     &12     \\
50       &0.33       &23.1       &0.53       &6.82      &30    &0.44        &13.6         &59    &1.04        &2.6     &11     \\
55       &0.46       &17.0       &0.74       &4.43      &26    &0.62        &9.1          &53    &1.52        &1.7     &10     \\
60       &0.71       &11.6       &1.15       &2.61      &23    &0.95        &5.5          &47    &2.33        &1.0     & 8.7   \\
63       &0.97       & 8.4       &1.60       &1.66      &20    &1.31        &3.6          &43    &3.19        &0.67    & 7.9   \\
65       &1.22       & 6.8       &2.03       &1.16      &17    &1.68        &2.6          &37    &4.09        &0.47    & 6.9   \\
67       &1.58       & 5.1       &2.65       &0.74      &15    &2.18        &1.7          &33    &5.39        &0.30    & 5.8   \\
69       &2.15       & 3.8       &3.64       &0.43      &11    &3.08        &1.1          &28    &7.15        &0.18    & 4.6   \\
70       &2.53       & 3.2       &4.20       &0.30      & 9    &3.39        &0.8          &24    &8.39        &0.11    & 3.6   \\
\hline                                                                                    
\end{tabular}                                                                             
\caption{\gr\ rate predictions from simulations for the standard, \emph{hard}
 and \emph{loose} selection cuts. This table is valid for a source with an energy spectrum similar to the Crab, 
 for observations at an offset of 0.5\dg\ (the usual observing mode). The cut selection efficiencies
 and (peak rate) energy thresholds in each case are also given.}
\label{tab:cuteff}
\end{table*}                                                                               

In order to simplify the application of the optical correction
discussed in section \ref{sec:opcorr}, this correction is not applied
in estimation of the effective area for each event. Since the
distribution of Cherenkov light scales with the shower energy to a
good approximation, the detection probability and hence the effective
area depends primarily on the amount of light arriving at the camera,
and not on the absolute energy of the \gr. Thus it is not necessary to
recalculate the effective area lookup tables when the actual optical
efficiency changes by a small amount. For larger changes this
correction breaks down due to effects of the system trigger on events
near the energy threshold.  The selection cuts are made using the
image intensity without optical correction, and the corrected
intensity is only applied in the energy estimation.  This method has
been tested on Monte Carlo simulated sources with reduced optical
efficiency, and it was verified that the correct flux is
reconstructed.

\begin{figure}
  \centering
  \includegraphics[width=0.5\textwidth]{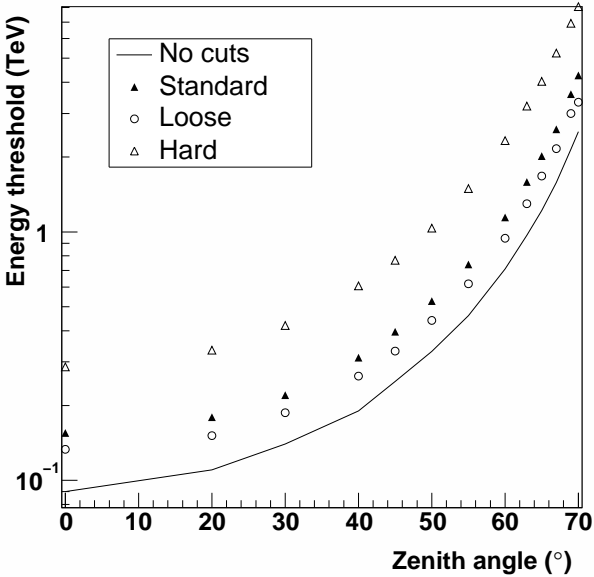}
  \caption{ The peak-rate energy threshold before and after selection cuts
  versus zenith angle, for three sets of selection cuts as described
  in Table \ref{tab:selcuts}. The safe threshold for spectral analysis in
  each case is slightly higher.}
\label{fig:thresh}
\end{figure}

Instead of using the safe energy threshold as introduced above, the
energy threshold for a set of observations has also been commonly
defined as the peak in the differential rate vs.  energy curve
\citep{konopelko99a}. This is formed by folding the effective area
curve, as plotted in Figure \ref{fig:effareas}(a) with the expected
\gr\ flux from the source. This energy threshold is generally slightly
lower than the safe threshold defined above, which is designed to
ensure an accurate spectral reconstruction. There may even be a
significant \gr\ signal below the safe threshold. The vertical lines
in Figure \ref{fig:effareas} define the safe energy thresholds for
each zenith angle.  Figure \ref{fig:thresh} shows the predicted
peak-rate energy threshold and \gr\ rate for a Crab-like source, based
on simulations and projection from the Crab flux as measured by the
HEGRA collaboration \citep{aharonian00b}.  Table \ref{tab:cuteff}
gives the pre-cut energy threshold as a function of zenith angle,
along with the equivalent after the various selection cuts. It should
be noted that the energy threshold, by this definition, depends on the
spectrum of the source.

The effective area also varies with the position of the source in the
field of view of the instrument. As larger energy showers are
preferentially detected at higher impact distances, they appear closer
to the edge of the field of view. Thus truncation of images for
sources closer to the edge of the field of view tends to reject events
at higher energies. Monte Carlo simulations are made at a range of
source offset positions in order to make effective area curves and the
effective area is interpolated for a particular observation. Table
\ref{tab:cuteff} also shows the event rates before selection cuts and
after each of the selection cuts described above, as a function of
zenith angle for a source similar in flux to the Crab nebula. These
predictions are based on the effective areas estimated for a source
offset by 0.5\dg\ from the observation position. The cut efficiencies
in each case are also shown, with the (peak rate) energy thresholds.

\section{Flux and spectral measurements}

The data from observations of the Crab have been analysed using the
technique described, individually by data set as outlined in Table
\ref{tab:obs}, and combined. The analysis has been carried out for
each of the sets of selection cuts as outlined in Table
\ref{tab:selcuts}, using the reflected background method with 5
background regions. Table \ref{tab:crabstat} outlines the numbers of
events passing cuts in the \on\ and \off\ regions, as well as the
background normalisation ($\alpha$), the number of excess events, the
significance ($\sigma$) of the excess, the rate of \grs\ passing cuts
and the \sigrate. Data Sets I-III are combined to give a total result.
For example, the mean rate of \grs\ for the standard selection cuts is 6.0 \pmin\,
with a significance of 27 \sigrate. It can be seen that the rate of
events passing cuts is strongly dependent on the zenith angle of the
observations, as well as on the selection cuts used. Comparisons
between Table \ref{tab:crabstat} and Table \ref{tab:cuteff} show that
the effective area estimation correctly reproduces the \gr\ rate for
the various data sets (within statistical errors).  The mean rate for
data set III, at a mean zenith angle of 54\dg, is predicted to be 4.8
\pmin, while the measured value is $(4.9 \pm 0.1\stat) \pmin$.  The
loose cuts at 55\dg\ zenith angle keep 53\% of the possible \grs,
making them more suitable for spectral studies of strong sources.

\begin{table*}
\centering
  \begin{tabular}{cccccccccccc}
\hline
Data Set  &method   &\on   &\off  &$\alpha$ &excess &significance &rate &\sigrate &$\textrm{F}_{>1 TeV}$&\rchisq &r.m.s. \\  
         &         &      &      &         &       &$\sigma$     &$\pmin$ &   &($\times 10^{-11} \iflux$)  &   & \%    \\
\hline                                                                                                                   
I       &std      & 1866 &  749 & 0.20 &  1718 & 62.2 &$  5.93 \pm 0.10 $ & 28.3  &$1.94 \pm 0.05$ & 14 / 10 & 11.0\\
II      &std      & 1976 & 1579 & 0.20 &  1667 & 53.2 &$  4.85 \pm 0.09 $ & 22.2  &$2.37 \pm 0.07$ & 27 / 13 & 16.0\\
III     &std      & 4759 & 2417 & 0.20 &  4283 & 94.2 &$  6.70 \pm 0.07 $ & 28.9  &$2.21 \pm 0.04$ & 53 / 25 & 12.1\\
\hline                                                                                                            
all     & std     & 8601 & 4745 & 0.20 & 7666  & 124  &$  6.0  \pm 0.05 $ & 27.0  &$2.16 \pm 0.03$ & 133 / 51 & 14.9\\
all     & loose   &27970 &61740 & 0.20 &15570  & 106  &$ 12.2  \pm 0.16 $ & 23.1  &$2.08 \pm 0.02$ & 143 / 51 & 22.2\\
all     & hard    & 3058 &  376 & 0.19 & 2986  &  94  &$  2.35 \pm 0.02 $ & 20.5  &$2.43 \pm 0.05$ &  87 / 51 & 20.0\\
all     & extended&25490 &24160 & 0.51 &13140  &  80  &$ 10.3  \pm 0.13 $ & 17.4  &$2.21 \pm 0.03$ & 128 / 51 & 22.2\\
all     & std Ring& 8525 & 6573 & 0.14 & 7588  & 129  &$  5.97 \pm 0.05 $ & 28.0  &$2.17 \pm 0.03$ & 133 / 51 & 15.0\\
\hline           

\end{tabular}
\caption{Events passing cuts in \on\ and \off\ regions for the Crab, listed by
  data set along with excesses, significance and \gr\ rates. Various
  selection cuts described in Table \ref{tab:selcuts} are also compared
  for data sets I-III combined. The results using the ring-background
  model are given (denoted as Ring). The integrated flux
  from the Crab above 1 TeV is shown also, as described in section
  \ref{sec:flux}. The \rchisq\ for a fit to a constant flux for
  the data set is given, as is the percentage run-by-run rms deviation in the
  flux.}
\label{tab:crabstat}
\end{table*}

\begin{figure*}
  \centering
  $\begin{array}{cccc}
    
    \multicolumn{1}{l}{\mbox{\bf }} &
    \multicolumn{1}{l}{\mbox{\bf }} \\ [0cm]
    \resizebox{0.33\hsize}{!}{\includegraphics[width=0.3\textwidth,height=10cm]{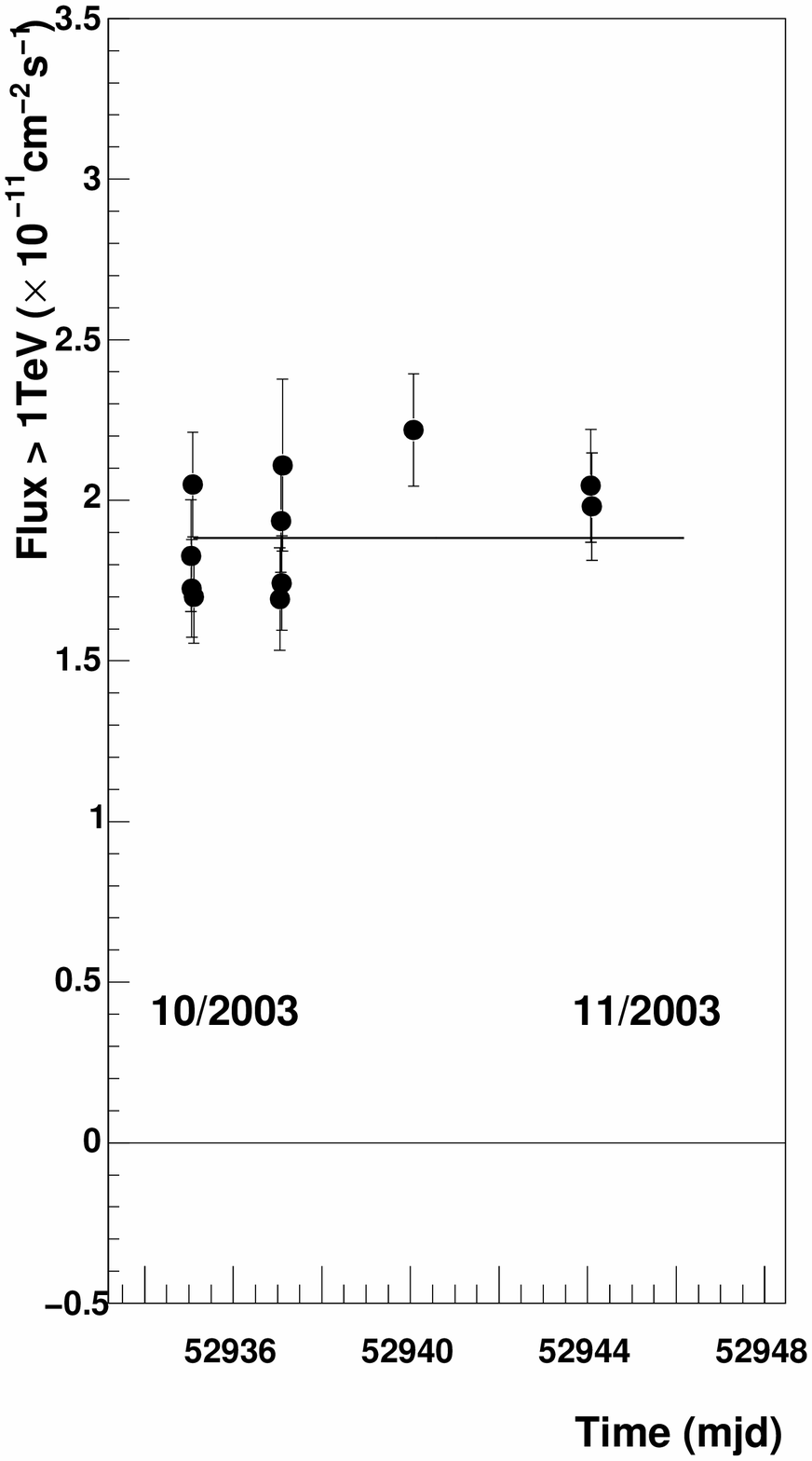}} &
    \resizebox{0.33\hsize}{!}{\includegraphics[width=0.3\textwidth,height=10cm]{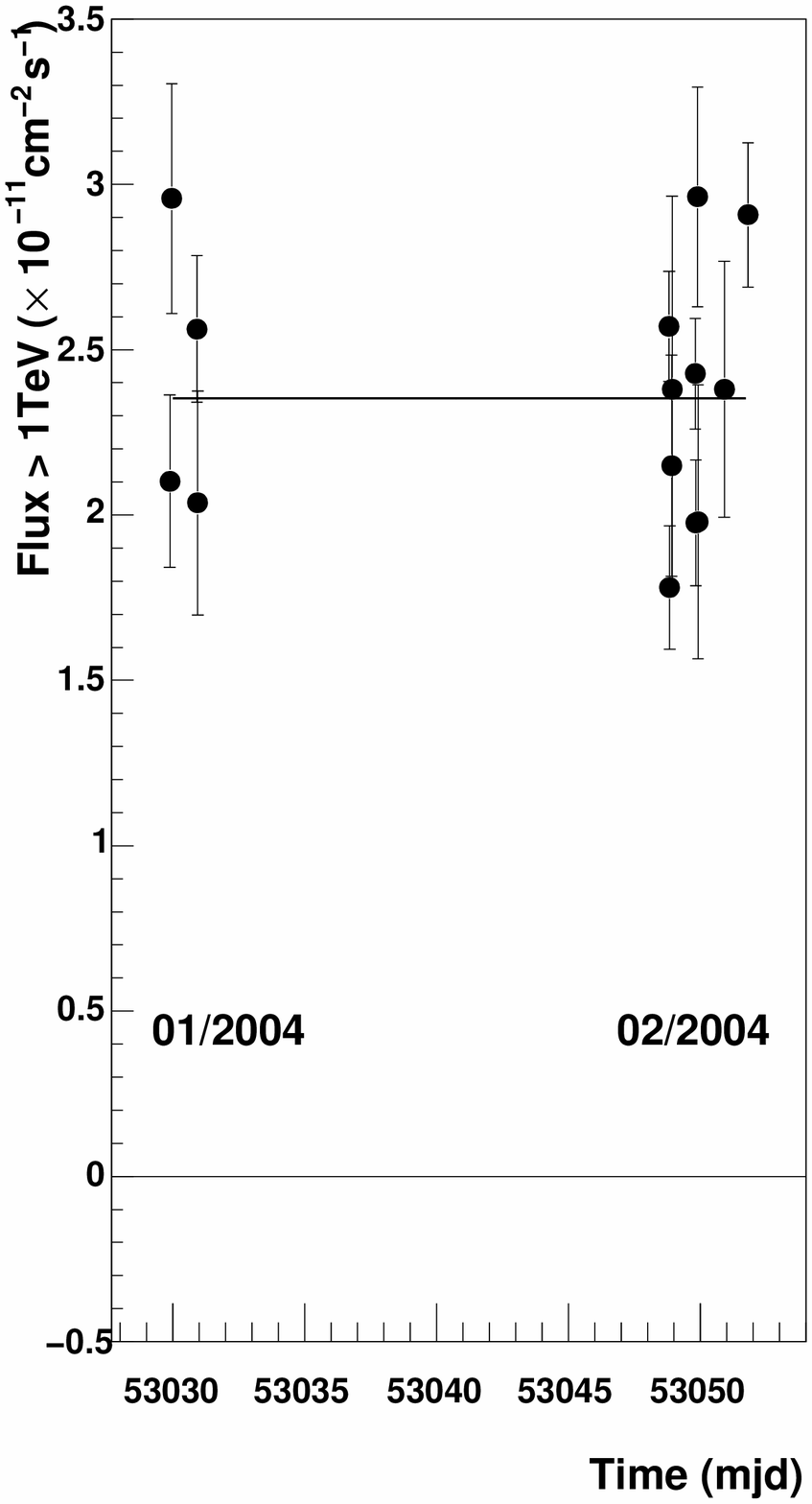}} &
    \resizebox{0.33\hsize}{!}{\includegraphics[width=0.3\textwidth,height=10cm]{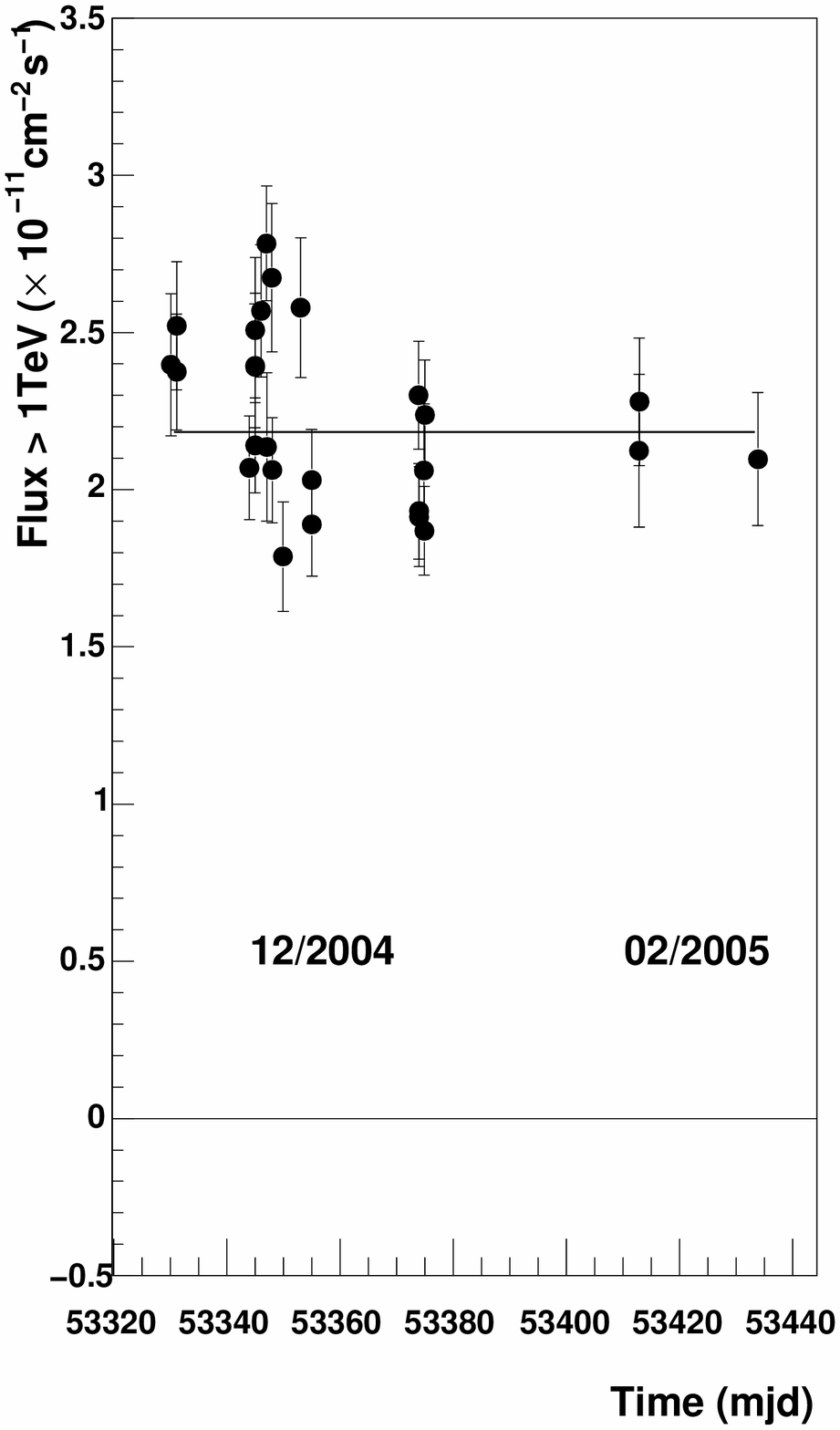}} \\ [0cm]
    \mbox{\bf (a)} & \mbox{\bf (b)}  & \mbox{\bf (c)}
  \end{array}$

  \caption{The run-by-run light curve of the integral flux above 1
    TeV for {\bf a)} data set I {\bf b)} data set
    II {\bf c)} data set III. All efficiency
    corrections as discussed in the text have been applied to these
    data.}
  \label{fig:lc}
\end{figure*}

\subsection{Run by run flux measurements}
\label{sec:flux}

For the purpose of producing a light curve of the \gr\ flux from a
source, the integrated flux above the threshold energy is calculated
for each time period ($t_{start}$ to $t_{stop}$), assuming a
particular spectral form for the source, such as a power law with
photon index $\Gamma$ and flux normalisation $\textrm{I}_{0}$ in units
of \dflux.  The excess number of events seen from a source ($\delta$)
is given by the following:

\begin{equation}
  \label{eq:binflux}
\delta = \int_{0}^{\mathrm{E}_{c}} \int_{t_{start}}^{t_{stop}} \textrm{I}_{0} (\frac{\mathrm{E}}{\mathrm{E}_{0}})^{-\Gamma} \
\atrue(E,Z(t))\ dt\ dE
\end{equation}

Here \atrue\ is the effective area as a function of zenith angle Z and
true energy $E$ from Monte Carlo simulations. The flux normalisation
can then be calculated by integrating the effective area up to some
upper cutoff energy $\mathrm{E}_{c}$ and over the integration time. The value
of $\mathrm{E}_{c}$ is imposed by the range of the Monte Carlo simulations and
is normally above 100 TeV. The integral flux above threshold is
usually quoted above the threshold energy for the observations, or
alternatively above 1 TeV. Here the latter convention is used for
simplicity.

\begin{figure}
  \centering
  \includegraphics[width=0.45\textwidth]{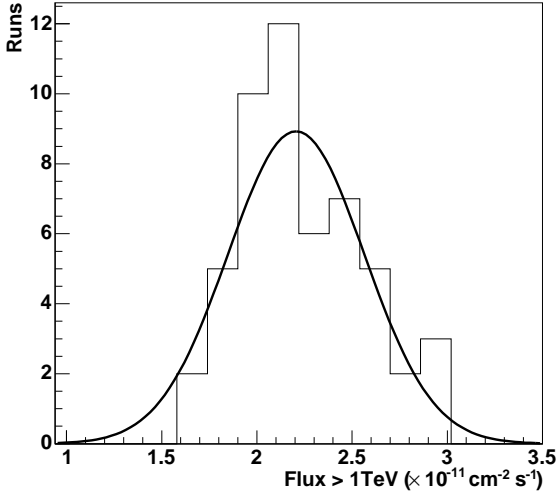}
  \caption{Distribution of run by run fluxes for data sets
  I-III. The fitted Gaussian distribution has a mean of $(2.21 \pm 0.06\stat) \times
  10^{-11} \iflux$ and a $\sigma$ of $(3.58 \pm 0.6\stat) \times 10^{-12}\iflux$.}
\label{fig:lcdist}
\end{figure}

The fluxes measured for each data set included in this analysis, along
with the mean fluxes for data sets I-III are given in Table
\ref{tab:crabstat}. The mean flux is also given for the various sets
of selection cuts described in Table \ref{tab:selcuts}. The rms
variation in the mean flux per data set is 8\%, while the rms
variation in the run-by-run fluxes is 14.9\%, the typical statistical
error on a single run is 5\% for moderate zenith angles and offsets.
Figure \ref{fig:lc} shows light-curves for data sets I, II and III,
while the distribution of run-by-run flux is shown in Figure
\ref{fig:lcdist}, for all of the data.

It can be seen that the long-term variations in the run-by-run flux
after the correction for changes in the detector optical efficiency
are small compared to the short-term variations, mainly due to
atmospheric effects. The mean flux in data sets II and III, taken in
January 2004, is 10\% higher than that seen in data set I, which was
taken in October of 2003. This difference is smaller than the rms
spread of either data set, and can be explained by differences in the
atmospheric conditions between the two periods. No correction
  is made to the run-by-run flux for short term variations in the
  atmospheric conditions. Such corrections are under study and will be
  the focus of a future paper.

Systematic errors in the flux due to selection cuts and effective area
estimates have been studied by applying the various selection cuts
described to the data and measuring the integral flux. The
results are outlined in Table \ref{tab:crabstat}. The rms in the
measured fluxes is 15\%. A possible systematic error in the integral
flux measurement due to the background estimation method has been
tested by applying the ring-background method, as described above to
calculate a flux. It can be seen that the reconstructed flux differs
only slightly with the two methods.

\subsection{Energy spectrum}

The energy spectrum of the Crab has been measured using the data
described here. The method used for deriving the energy spectrum is
similar to those described by \citet{mohanty98}, \citet{aharonian99b}
and \citet{aharonian04a}. An energy spectrum is fit for each data set
and for the combined data.

The bin size for the energy spectrum is set depending on the overall
significance of the signal. The maximum possible energy bin is defined
by the simulations, which extend to above 400 TeV at 50\dg\ zenith
angle.  Only energies above the safe threshold as defined above are
used for spectral determination.

In each energy bin $i$ above the minimum energy the differential flux
is calculated by summing over the on source events $N_{on}$, weighted
by the inverse of the effective area (\areco) as a function of the
reconstructed energy of each event. The normalised sum of the weighted
off events $N_{off}$ is then subtracted. The difference is weighted by the
live-time for that bin (T) and the bin width ($\Delta
\mathrm{E}_{i}$):

\begin{equation}
  \label{eq:intflux}
  \frac{dF_{i}}{dE} =
  (\mathrm{T}\,\Delta \mathrm{E}_{i})^{-1} . \
  \left(\sum_{j=0}^{Non}{(\mathrm{A}reco_{j})^{-1}} - \alpha
    \sum_{k=0}^{Noff}{(\mathrm{A}reco_{k})^{-1}}\right)
\end{equation}

In the case where runs are combined with varying zenith angles, and
thus varying useful energy thresholds, the live-time is calculated for
each energy bin separately. The result is scaled by the appropriate
live-time to give the \gr\ flux in each energy bin. The error on the
flux in each energy bin is estimated using standard error propagation.
A spectral energy function (for example a power law distribution) is
fit to the flux points using the least-squares method.  The maximum
energy for spectral fitting is chosen so as to have a significance in
that bin greater than 2$\sigma$.

\begin{figure}
  \centering
  \includegraphics[width=0.45\textwidth]{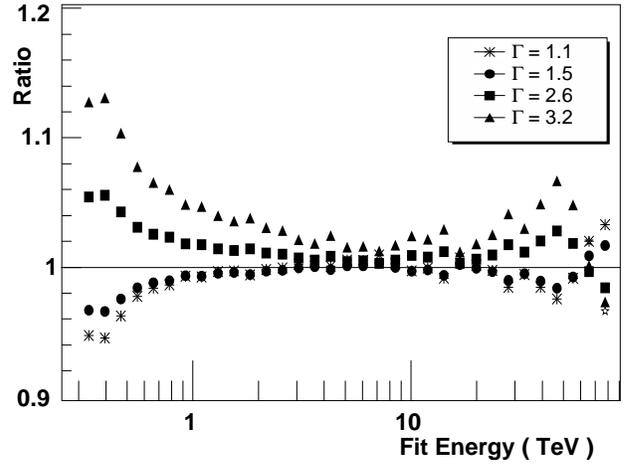}
  \caption{The ratio of the reconstructed effective area (estimated assuming an
    photon index of 2.0) to the true effective area per energy bin,
    for true photon indices from 1.1 to 3.2, based on Monte Carlo
    simulations at 45\dg\ zenith angle.}
\label{fig:specbias}
\end{figure}

Since the estimation of the effective area as a function of
reconstructed energy, \areco, depends on an assumed spectral slope, it
is strictly correct to adjust \areco\ for the fitted spectrum and then
re-fit, repeating until the fit converges \citep{aharonian99b}. Figure
\ref{fig:specbias} shows the bias as a function of energy introduced
by using an assumed photon index of 2.0 for the effective area
estimation, given true spectra with indices ranging from 1.1 to 3.2.
It can be seen that at 440 GeV, the differential flux for a source
with an intrinsic photon index of 2.6 is overestimated by 5\%, while
the differential flux given a true photon index of 1.5 is
underestimated by 4\%. For energies well above threshold the bias is
less than 5\% for a wide range of photon indices. Thus the effect of
this correction on the Crab spectrum is small and was neglected for
this analysis.

\begin{figure*}
  \centering
  $\begin{array}{cc}
    \multicolumn{1}{l}{\mbox{\bf }} &
    \multicolumn{1}{l}{\mbox{\bf }} \\ [0cm]
    \resizebox{0.445\hsize}{!}{\includegraphics[width=0.47\textwidth]{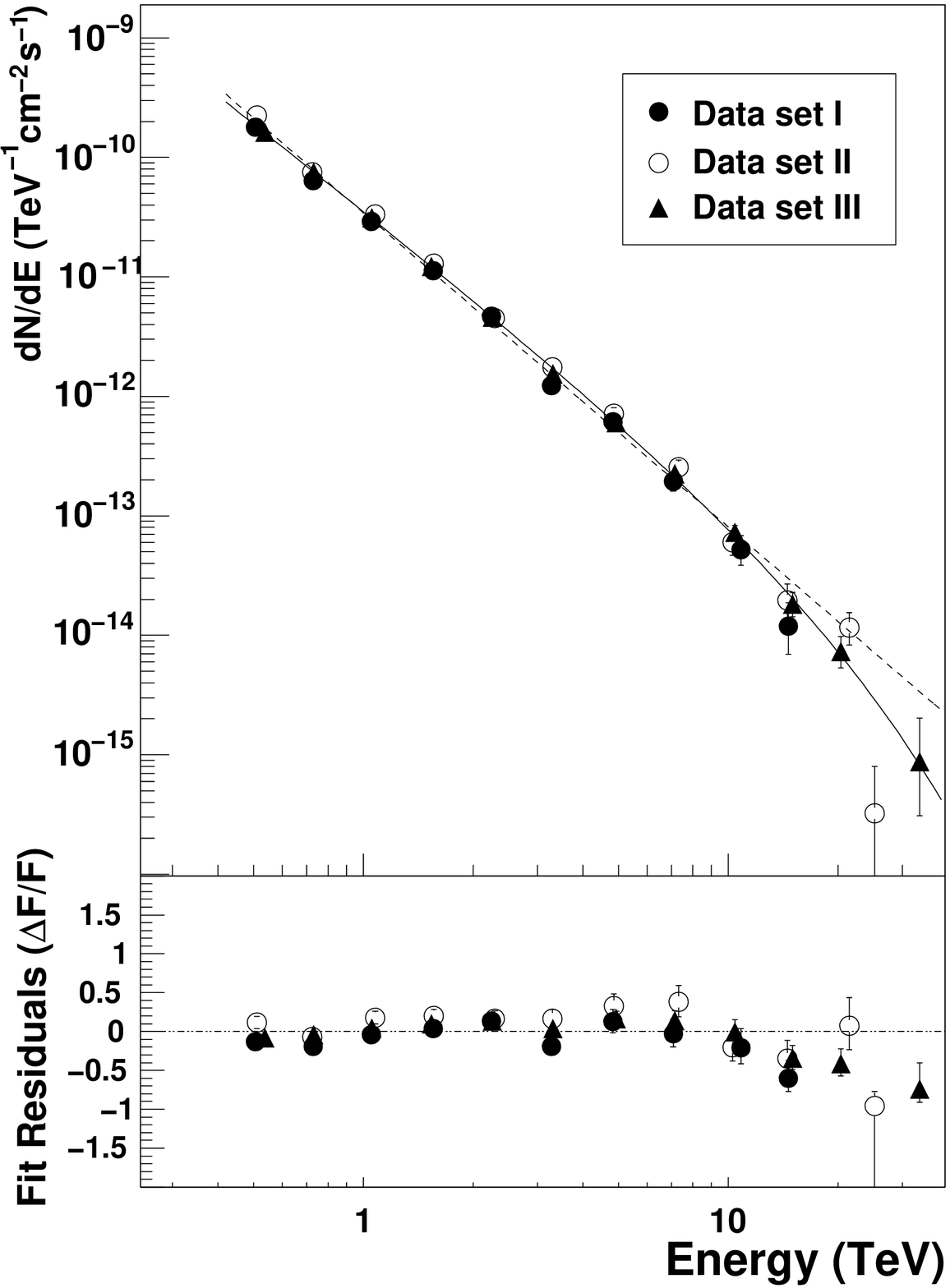}}&
    \resizebox{0.46\hsize}{!}{\includegraphics[width=0.47\textwidth]{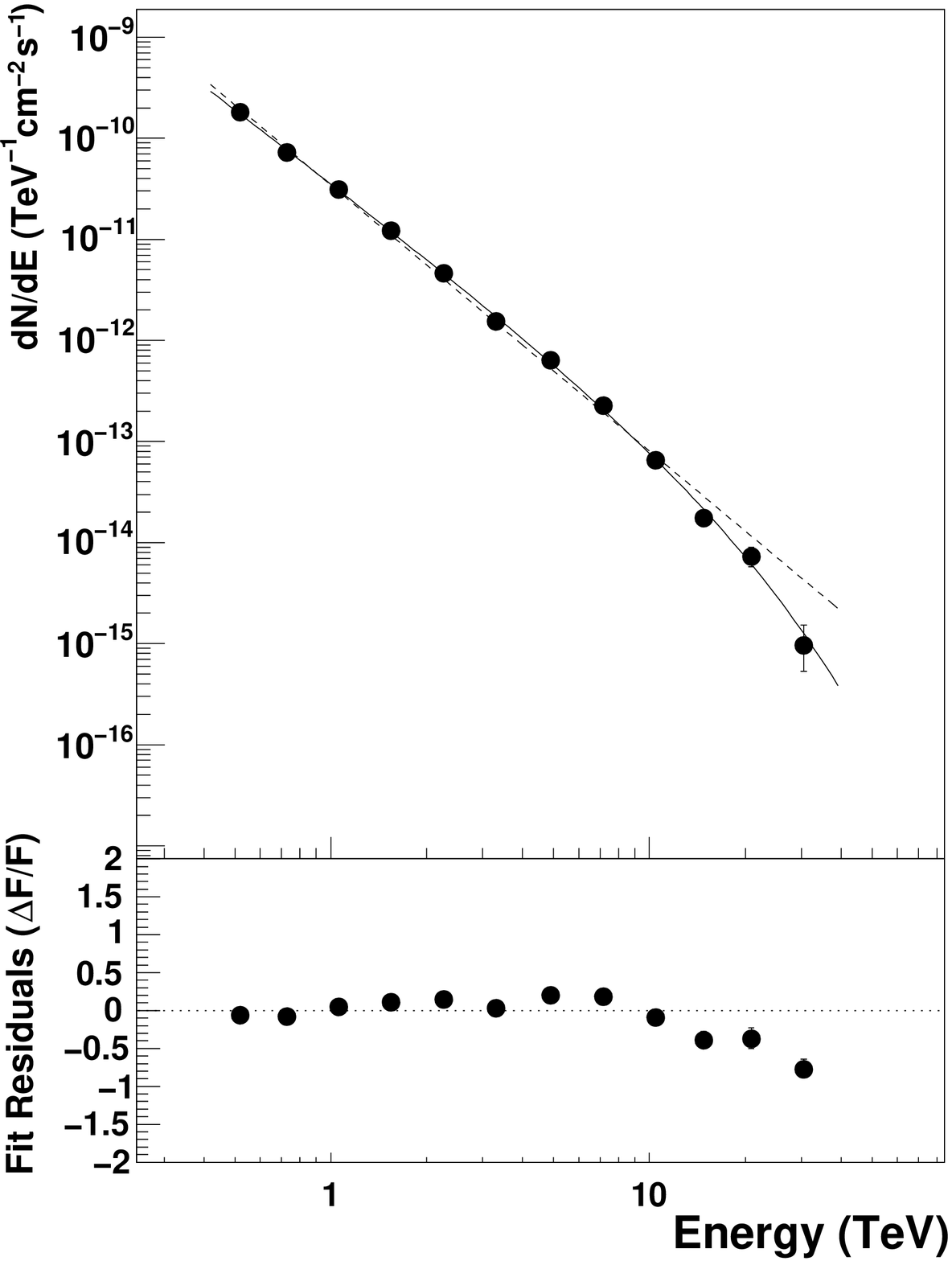}}\\ [0cm]
    \mbox{\bf (a)} & \mbox{\bf (b)}                                
  \end{array}$

  \caption{{\bf a)}  Energy spectra
    for data sets I (filled circle), II (open circles) and III (filled
    triangle), fit residuals to the common power law fit are also
    shown.  The dashed line indicates the best fit power law spectrum,
    while the solid line denotes the fit including an exponential
    cutoff. {\bf b)} Combined average energy spectrum for data sets
    I-III, fit residuals to the combined power law fit are shown.}
  \label{fig:spec}
\end{figure*}

\begin{table*}
\centering
  \begin{tabular}{lccc}
\hline
Mean energy& excess events  & Significance & Differential flux $\left(\frac{\mathrm{dN}}{\mathrm{dE}}\right)$\\
(TeV)      &                & $\sigma$     & (\dflux)              \\
\hline
0.519   & 975           & 42.9         &$(1.81 + 0.06 - 0.06) \times 10^{-10}$\\
0.729   & 1580          & 56.0         &$(7.27 + 0.20 - 0.19) \times 10^{-11}$\\
1.06    & 1414          & 55.3         &$(3.12 + 0.09 - 0.09) \times 10^{-11}$\\
1.55    & 1082          & 47.3         &$(1.22 + 0.04 - 0.04) \times 10^{-11}$\\
2.26    & 762           & 39.5         &$(4.6  + 0.18 - 0.18) \times 10^{-12}$\\
3.3     & 443           & 29.5         &$(1.53 + 0.08 - 0.08) \times 10^{-12}$\\
4.89    & 311           & 24.9         &$(6.35 + 0.39 - 0.38) \times 10^{-13}$\\
7.18    & 186           & 19.6         &$(2.27 + 0.18 - 0.17) \times 10^{-13}$\\
10.4    & 86            & 13.1         &$(6.49 + 0.77 - 0.72) \times 10^{-14}$\\
14.8    & 36            &  8.1         &$(1.75 + 0.33 - 0.30) \times 10^{-14}$\\
20.9    & 23            &  7.5         &$(7.26 + 1.7  - 1.50) \times 10^{-15}$\\
30.5    & 4             &  2.9         &$(9.58 + 5.6  - 4.25) \times 10^{-16}$\\
\hline                                     
\end{tabular}
\caption{Flux measurements for each energy bin in the combined spectral fit on data sets I-III,
 as plotted in Figure \ref{fig:spec}(b). The flux errors are error-propagated 68\% Feldman-Cousins confidence
 intervals \citep{feldman98}.}
\label{tab:crabpoints}
\end{table*}

The number of excess events and significance is given for each energy
bin along with the differential flux in Table \ref{tab:crabpoints}.
Only statistical errors are given here. A significant signal is seen
in every energy bin from the threshold energy of 440 GeV up to 20 TeV,
and a marginal signal is seen at the maximum bin of mean energy 30.5
TeV. The spectrum is shown in figure \ref{fig:spec}. The fit of a
power law function to the combined data with the standard analysis
cuts yields $\Gamma = 2.63 \pm 0.02\stat$ and differential flux
normalisation at 1 TeV $\mathrm{I}_{0} = (3.45 \pm 0.05\stat)\ \times
10^{-11} \dflux$ (PL in Table \ref{tab:crabspec}).

In the combined spectral fit there is evidence for a steepening of the
energy spectrum; a fit of a power law with an exponential cutoff:
$\frac{\mathrm{dN}}{\mathrm{dE}} = \mathrm{I}_{0}\ (\mathrm{E}/1 \mathrm{TeV})^{(-\Gamma)}
\exp(-\mathrm{E}/\mathrm{E}_{c})$, gives a differential flux normalisation at 1 TeV $\mathrm{I}_{0} = (3.76 \pm
0.07\stat) \times 10^{-11} \dflux$, with $\Gamma = 2.39 \pm 0.03\stat$ and a
cutoff energy $\mathrm{E}_{c} = (14.3 \pm 2.1\stat)$ TeV. The \chisq\ for this
fit is 15.9 with nine degrees of freedom. This compares with a \chisq\ 
of 104 with ten degrees of freedom for the straight power law fit,
thus the fit including an exponential cutoff is clearly favoured. A
broken power law fit  (BPL in Table \ref{tab:crabspec}): 

$\frac{\mathrm{dN}}{\mathrm{dE}} = \mathrm{I}_{0}\ \left(\frac{\mathrm{E}}{\mathrm{E}_{c}}\right)^{-\Gamma_{1}}
\left( 1 + \left(\frac{\mathrm{E}}{\mathrm{E}_{c}}\right)^{1/\mathrm{S}}\right)^{\mathrm{S}(\Gamma_{1}-\Gamma_{2})}$

gives a differential flux normalisation $\mathrm{I}_{0} = (3.43 \pm
0.07\stat) \times 10^{-11} \dflux$, with $\Gamma = 2.51 \pm 0.02\stat$
below the break energy of $\mathrm{E}_{c} = (7.0 \pm 0.1\stat)$ TeV. The photon
index above the break is $\Gamma = 3.3 \pm 1.5\stat$, and the \rchisq\ 
of the broken power law fit is 28.6/8. This fit includes a term (S) for the width of
the transition region, which is fixed to 0.3.

\begin{table*}
\centering
  \begin{tabular}{llllllllll}
\hline
Data Set &Selection &$\mathrm{E}_{min}$ &$\mathrm{E}_{max}$&$\mathrm{I}_{0} (1 \mathrm{TeV})$  &$\Gamma$  &$\mathrm{E}_{c}$&$\rchisq$  &$\mathrm{F}_{>1 TeV}$\\  
          &cuts      &(TeV)     &(TeV)    &($\times 10^{-11} \dflux$)&          &(TeV) &           &($\times 10^{-11} \iflux$)\\
\hline    
I         & std     &0.41      & 19      &$ 3.53 \pm 0.17$    &$2.37 \pm 0.07$ &$ 11.2 \pm   4.2$& 11.8 / 7 &$2.06 \pm 0.20$\\
II        & std     &0.41      &100      &$ 4.36 \pm 0.16$    &$2.30 \pm 0.06$ &$  8.4 \pm   1.2$& 26.3 /10 &$2.48 \pm 0.16$\\
III       & std     &0.45      & 65      &$ 3.84 \pm 0.09$    &$2.41 \pm 0.04$ &$ 15.1 \pm   2.8$& 12.6 / 9 &$2.31 \pm 0.10$\\
\hline        
all      & std      &0.41      & 40      &$ 3.76 \pm 0.07$    &$2.39 \pm 0.03$ &$ 14.3 \pm   2.1$& 15.9 / 9 &$2.26 \pm 0.08$\\
all (BPL)& std      &0.41      & 40      &$ 3.43 \pm 0.07$    &$2.51 \pm 0.02\ (3.3 \pm 1.5)$ &$  7.0 \pm   0.1$& 28.6 / 8 &$2.24 \pm 0.06$\\
all (PL) & std      &0.41      & 40      &$ 3.45 \pm 0.05$    &$2.63 \pm 0.01$ &                 &104  / 10 &$2.11 \pm 0.03$\\
all      & loose    &0.34      & 71      &$ 3.53 \pm 0.06$    &$2.37 \pm 0.03$ &$ 17.6 \pm   2.6$&  7.1 /10 &$2.22 \pm 0.07$\\
all      & hard     &0.73      & 71      &$ 4.06 \pm 0.12$    &$2.53 \pm 0.05$ &$ 20.3 \pm   4.5$& 17.4 / 8 &$2.36 \pm 0.12$\\
all      & extended &0.45      & 30      &$ 3.78 \pm 0.07$    &$2.30 \pm 0.04$ &$ 14.8 \pm   2.5$& 14.6 / 8 &$2.41 \pm 0.10$\\
\hline
all(uncorr.)& std      &0.34      & 33      &$ 3.02 \pm 0.06$    &$2.45 \pm 0.03$ &$ 13.7 \pm   2.0$& 21.8 / 9 &$1.75 \pm 0.06$\\
all(ring) & std      &0.41      & 86      &$ 3.76 \pm 0.07$    &$2.40 \pm 0.03$ &$ 15.1 \pm   2.4$& 14.9 /10 &$2.27 \pm 0.08$\\
\hline
Whipple   &          &          &         &$ 3.20 \pm 0.17$    &$2.49 \pm 0.06$ &                 &          &$2.1  \pm 0.2$\\
CAT       &          &          &         &$ 2.20 \pm 0.05$    &$2.80 \pm 0.03$ &                 &          &$1.22 \pm 0.03$\\
HEGRA     &          &          &         &$ 2.83 \pm 0.04$    &$2.62 \pm 0.02$ &                 &          &$1.75 \pm 0.03$\\  
\hline           
\end{tabular}

\caption{Flux and spectral measurements of the Crab, divided up by data set as outlined in Table \ref{tab:obs},
 for a power law fit with an exponential cutoff. 
Results for the various selection cuts described in Table \ref{tab:selcuts} are also compared. The results for
 a power-law fit (PL) and for a broken power-law fit (BPL) are also given. The spectral fit estimated using the ring
 background model is given (ring), as is that estimated without the optical efficiency correction (uncorr.).
 Only statistical results are shown in the table.  Similar measurements from other experiments are given for comparison,
 the Whipple results is taken from \citet{mohanty98}, the CAT results from \citet{masterson99} and the HEGRA results from \citet{aharonian04a}.}
\label{tab:crabspec}
\end{table*}

The flux and spectral measurements for the separate data sets are
summarised in Table \ref{tab:crabspec}. The rms spread of the photon
index from data set to data set is 0.04. The rms spread in the
integral flux, calculated from the fitted spectrum, is 15\%, with a
statistical error of 2\% on the integrated flux for the combined data.
Figure \ref{fig:spec}(a) shows the energy spectral points superimposed
for the three data sets used in the spectral analysis, the residuals
about the combined fitted spectrum are shown underneath. Figure
\ref{fig:spec}(b) shows the energy spectral fit, along with residuals,
for the combined data sets I-III. The spectral fits for the various
selection cuts are also included in Table \ref{tab:crabspec}, as is
the fit for the ring background model. For comparison the fit spectrum
calculated without the optical efficiency correction is also included.

\begin{figure}
  \centering
  \includegraphics[width=0.5\textwidth]{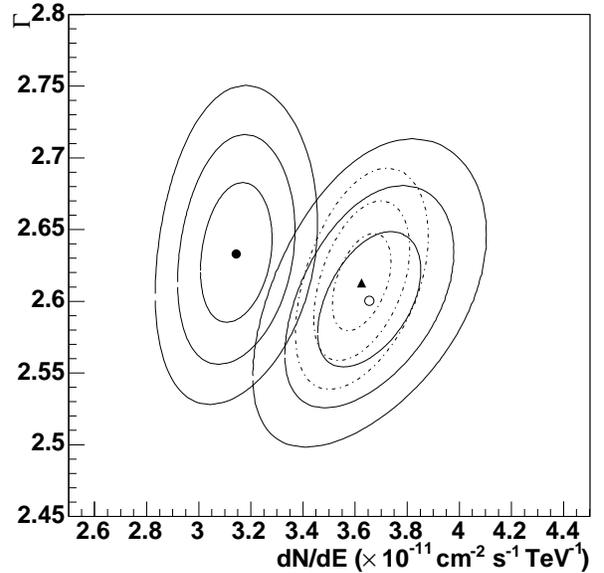}
  \caption{Contour plot of the \chisq\ fit error as a function of the
    power law parameters F$_{0}$ and $\Gamma$ for data sets I (filled
    circle), II (open circle) and III (filled triangle, dotted lines). The 68\%,
    95\% 99.9\% error contours are shown, and The best fit spectral
    parameters are marked in each case.}
\label{fig:speccont}
\end{figure}

Figure \ref{fig:speccont} shows a 2-d plot of the fitted photon index
$\Gamma$ against the flux normalisation for each data set analysed,
here a simple power law fit has been made for simplicity. The error
contours, estimated using the least-squares method, are also shown. It
can be seen that the three data sets are compatible at the $2\sigma$
level; data set II includes large zenith angle data, and is fitted
with a slightly softer spectrum, caused by the higher energy threshold
of these observations and the significant curvature seen in the
spectral measurements.

The combined data sets I-III have also been analysed with the various
selection cuts described in Table \ref{tab:selcuts}, and a spectrum
fitted. The rms spread in the photon index between the various
analyses is 0.08, however this includes the effect of the very
different energy threshold for the \emph{hard} selection cuts, which
may give rise a softer photon index if the source spectrum is
intrinsically curved. The rms spread of the reconstructed integral
flux is 8\%, which indicates that the reconstructed flux is not
strongly dependent on the details of the analysis method. The use of
the ring-background method results in a flux and spectral slope
similar to that reconstructed using the standard reflection
method.

\subsection{Estimation of systematic errors}
The systematic error on the absolute flux is estimated from the
various independent contributing factors, as discussed in section
\ref{sec:systematics}. These errors are summarised in Table
\ref{tab:systematics}. The total estimated systematic error, after
correction for degradation in the optical efficiency, is 20\%.  The
sources of systematic error include uncertainties due to the shower
interaction model and the atmospheric model used in the Monte Carlo
simulations. Also included is the estimated uncertainty in the flux
due to the effect of missing pixels, which has been conservatively
estimated at 5\%, and the effect of uncertainty in the live-time
measurement, which is less than 1\%. The uncertainty in flux due to
selection cuts is estimated from the rms of the flux and spectral
slope measurements detailed in Table \ref{tab:crabspec}, as is the
uncertainty due to the background model.  The run-by-run rms over the
entire set of data is 15\%, this is thought to mainly be due to
variations in the atmosphere, and thus is included in the systematic
error as an independent factor. The rms of the spectral estimations
for the various datasets in Table \ref{tab:crabspec} is used to
estimate the uncertainty in the spectral slope, which is 0.1.

\section{Conclusions}

A strong signal has been detected from the Crab nebula during the
commissioning phase of the \hess\ instrument and with the complete
instrument. An energy spectrum has been measured, with a differential
spectrum described by a power law with slope $\Gamma = 2.39 \pm
0.03\stat \pm 0.09\sys$ and an exponential cutoff at $(14.3 \pm
2.1\stat \pm 2.8\sys)$ TeV. The integral flux above 1 TeV is $(2.26
\pm 0.08\stat \pm 0.45\sys) \times 10^{-11} \iflux$.

\begin{table}
\centering
  \begin{tabular}{ccc}
\hline
Uncertainty                        &Flux   & Index \\
\hline
MC Shower interactions             & 1\%   &       \\
MC Atmospheric sim.                & 10\%  &       \\
\hline                                             
Broken pixels                      & 5\%   &       \\
Live time                          & 1\%   &       \\
\hline                                             
Selection cuts                     & 8\%   & 0.08  \\
Background est.                    & 1\%   & 0.01  \\
Run-by-run variability             & 15\%  &  -    \\
Data set variability               &  -    & 0.05  \\
\hline
Total                              & 20\%  & 0.09  \\
\hline
\end{tabular}
\caption{Summary table showing the various estimated contributions to the systematic flux error}
\label{tab:systematics}
\end{table}

\begin{figure}
  \centering
  \includegraphics[width=0.48\textwidth]{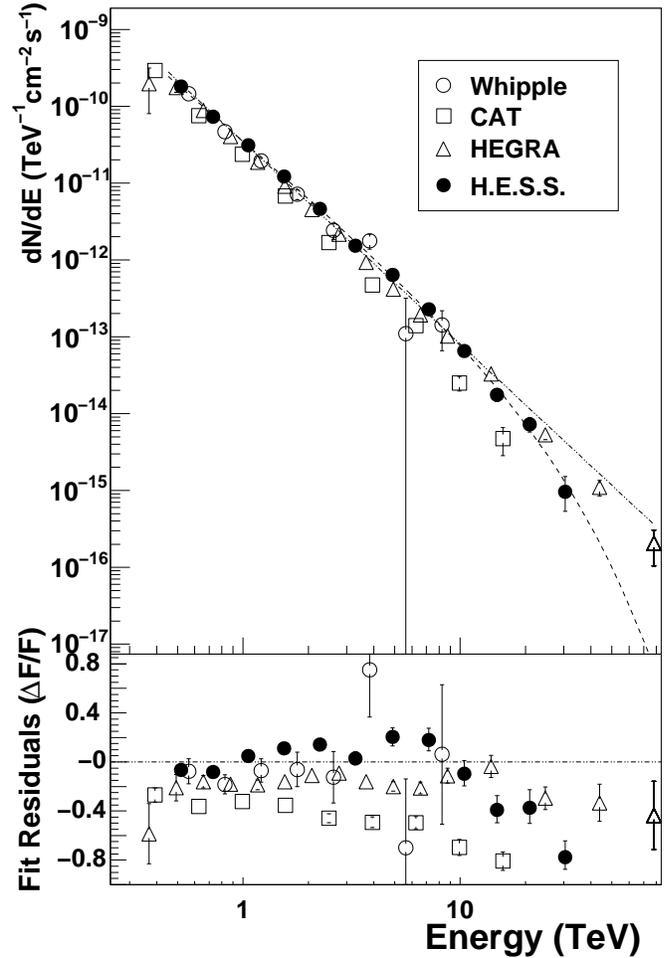}
  \caption{Comparison between the spectral measurements of CAT (open
    squares), Whipple (open circles) and HEGRA (open triangles) and
    the results if this study (filled circles). The residuals are
    shown for each spectrum relative to a power law fit to the \hess\ 
    data (dotted line).}
\label{fig:speccomp}
\end{figure}

Marginal steepening in the spectrum measured on the Crab nebula has
been previously claimed by \citet{aharonian04a} in studies of the Crab
with the HEGRA experiment.  Figure \ref{fig:speccomp} compares the
Crab spectrum from this study with measurements by HEGRA, CAT and
Whipple. Acceptable agreement is seen up to 10 TeV between the
experiments, although the CAT result gives a somewhat steeper
spectrum; the rms variation in integral flux between the four
experiments is 22\%.  Above 10 TeV the energy spectrum as seen by
\hess\ steepens significantly, in particular compared to HEGRA.

\begin{figure}
  \centering
  \includegraphics[width=0.5\textwidth]{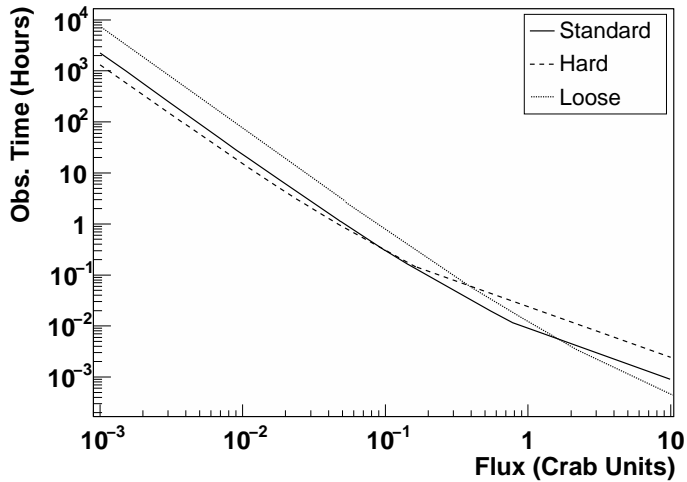}
  \caption{Sensitivity of the HESS array, expressed as the amount of
    time required to detect a signal at the $5\sigma$ level, as a
    function of the flux of the source, for a source of similar
    spectral slope to the Crab nebula (for observations at 20\dg\ 
    zenith angle, 0.5\dg\ offset from the source). A minimum of 10
    excess events is also required for very short observations. Shown
    are curves for the \textit{standard}, \textit{hard} and
    \textit{loose} selection cuts.}
\label{fig:sens}
\end{figure}

The softening seen in the Crab spectrum at high energies is consistent
with models of inverse Compton emission due to a population of
electrons extending up to PeV energies. Due to the high magnetic field in the
Crab nebula, the dominant target photon field for emission is probably
created by synchrotron emission from the same electron population
\citep{hillas98}.  More detailed models of \gr\ emission in the Crab
nebula are discussed by \citet{atoyan96, dejager92}.

Given the agreement between the Monte Carlo simulations and the data,
one can use the simulations to predict the time required to detect a
point source of a certain strength as a function of zenith angle.  In
Figure \ref{fig:sens} we show the time as a function of signal
strength required for a $5\sigma$ detection at 20\dg\ zenith angle,
for the selection cuts described here.  The \hess\ array is capable of
detecting a point source with a flux of 1\% of the Crab nebula in 25
hours, or alternatively detecting a source of similar strength to the
Crab in 30 seconds. The sensitivity for extended sources decreases
approximately linearly with the source extension. This sensitivity is
unprecedented in the field of VHE astrophysics and opens a new window
for sensitive and precise measurements of VHE \gr\ sources.

\begin{acknowledgements}
The support of the Namibian authorities and of the University of Namibia
in facilitating the construction and operation of H.E.S.S. is gratefully
acknowledged, as is the support by the German Ministry for Education and
Research (BMBF), the Max Planck Society, the French Ministry for Research,
the CNRS-IN2P3 and the Astroparticle Interdisciplinary Programme of the
CNRS, the U.K. Particle Physics and Astronomy Research Council (PPARC),
the IPNP of the Charles University, the South African Department of
Science and Technology and National Research Foundation, and by the
University of Namibia. We appreciate the excellent work of the technical
support staff in Berlin, Durham, Hamburg, Heidelberg, Palaiseau, Paris,
Saclay, and in Namibia in the construction and operation of the
equipment.
\end{acknowledgements}


\end{document}